\documentclass[onecolumn,floatfix,superscriptaddress,showpacs,showkeys,nofootinbib]{revtex4}
\usepackage{epsfig}
\usepackage{amssymb,latexsym,amsmath}
\newcommand{\eq}[1]{\begin{align} #1 \end{align}}

\begin{document}
\title{ Particle Number
Fluctuations in Relativistic Bose and Fermi Gases}

\author{V.V. Begun}
\affiliation{
 Bogolyubov Institute for Theoretical Physics, Kiev, Ukraine}
\author{  M.I. Gorenstein}
\affiliation{
 Bogolyubov Institute for Theoretical Physics, Kiev, Ukraine}
\affiliation{Frankfurt Institute for Advanced Studies, Frankfurt, Germany}

\begin{abstract}
Particle number fluctuations are studied in relativistic Bose and
Fermi gases. The calculations are done within both the grand
canonical and canonical ensemble. The fluctuations in the
canonical ensemble are found to be different from those in the
grand canonical one. Effects of quantum statistics increase in 
the grand canonical ensemble for large chemical potential. This is,
however, not the case in the canonical ensemble. In the limit 
of large charge density a strongest difference between the grand canonical 
and canonical ensemble results is observed.

\end{abstract}


\maketitle

\section{Introduction}

% A difference between the
%number of particles and antiparticles (in what follows we use the
%names 'positively charged' and 'negatively charged' particles,
%respectively) corresponds to a conserved charge $Q=N_+-N_-$. Its
%average value, $\langle Q\rangle_{g.c.e.}$, is regulated by the
%chemical potential in the grand canonical ensemble (GCE), and the
%charge $Q$ is exactly fixed in the canonical ensembles (CE). The
%most strong Bose and Fermi effects for the fluctuations of $N_+$
%(or $N_-$) are observed in the GCE at large positive (or negative)
%charge densities. On the other hand, just in this limit we find
%the most dramatic differences between the GCE and CE results for
%the particle number fluctuations.

The statistical models have been successfully used to describe the
data on hadron multiplicities in relativistic nucleus-nucleus
(A+A) collisions (see, e.g., Ref.\,\cite{stat-model} and recent
review \cite{BMST}). This has stimulated an investigation of the
properties of  these statistical models. In particular, 
connections between different statistical ensembles for a
system of relativistic particles have been
intensively discussed. In A+A collisions one prefers to use the
grand canonical ensemble (GCE) because it is the most convenient
one from the technical point of view. The canonical ensemble (CE)
\cite{ce-a,ce,ce-b,ce-c,ce-d,ce-e} or even the microcanonical
ensemble (MCE) \cite{mce} have been used in order to describe the
$pp$, $p\bar{p}$ and $e^+e^-$ collisions when a small number of
secondary particles are produced.  At these conditions the
statistical systems are far away from the thermodynamic limit, so
that the statistical ensembles are not equivalent, and the exact
charge or both energy and charge conservation laws have to be
taken into account.
%The CE is relevant for the systems with a
%large number of all produced particles, e.g., the large number of
%created pions  or the number of nucleons in p+A collisions, but a
%small number ( of carriers of conserved
%charges like number of strange hadrons \cite{ce-c}, antibaryons
%\cite{ce-d}, or charmed hadrons \cite{ce-e}. This can happen not
%only in elementary $pp$, $p\bar{p}$, and $e^+e^-$, but also in p+A
%or even A+A collisions.
%
%
The CE suppression effects for particle multiplicities are well
known in the statistical approach to hadron production,
%and they are successfully applied to a statistical
%description of hadron production in high energy collisions
%\cite{ce-a,ce,ce-b,ce-c,ce-d,ce-e}. The CE formulation explains,
%for example,
e.g., the suppression in a production of strange hadrons
\cite{ce-c} and antibaryons \cite{ce-d} in small systems, i.e.,
when the total numbers of strange particles or antibaryons are
small (smaller than or equal to 1).
%This consideration
%demonstrates a difference of
The different statistical ensembles are not equivalent for small
systems. When the system volume increases, $V\rightarrow\infty$,
the average quantities in the GCE, CE and MCE become equal, i.e.,
all ensembles are thermodynamically equivalent.

The situation is different for the statistical fluctuations. 
The fluctuations in relativistic
systems are studied in event by event analysis of high
energy particle and nuclear collisions (see, e.g., Refs.~
\cite{fluc,step,step1,fluc1} and references therein).
%For a first time
%
In  the relativistic system of created particles,
%, relevant for the statistical description of
%hadron production in high energy collisions, the situation is
%different. 
only the net charge $Q=N_+-N_-$ (e.g., electric
charge, baryonic number, and strangeness) can be fixed. In the statistical
equilibrium an average
value of the net charge is fixed in the GCE, or exact one in the
CE, but $N_+$ and $N_-$ numbers fluctuate in both GCE and CE.
%These fluctuations are, however, very different in the GCE and CE.

The particle number fluctuations for the relativistic case in the
CE were calculated for the first time in Ref.~\cite{ce-fluc} for
the Boltzmann ideal gas with net charge equal to zero. These
results were then extended for the CE \cite{ce2-fluc,ce3-fluc,bec}
and MCE \cite{mce-fluc,mce2-fluc} and compared with the
corresponding results in the GCE (see also Ref.~\cite{turko}). The
particle number fluctuations have been found to be suppressed in
the CE and MCE in a comparison with the GCE.  This suppression survives
in the limit $V\rightarrow\infty$, so that the
thermodynamical equivalence of all statistical ensembles refers to
the average quantities, but does not apply to the fluctuations.

%In the textbooks of statistical mechanics discussion is usually
%limited to non-relativistic case, so that in the CE the particle
%number $N$ is fixed exactly and its fluctuations are absent. 

%In Ref.~\cite{ce2-fluc} the particle number fluctuations have been
%already studied for quantum gases in the GCE and CE. That
%consideration was, however, restricted  only to the case of zero
%conserved charge value (i.e., a zero value of the corresponding
%chemical potential). In the present paper At large charge
%densities the most strong Bose and Fermi effects for particle
%number fluctuations are observed in the GCE. However, the main
%result of the present paper is that just in the limit of large
%charge density there are the striking differences between the CE
%and GCE results for the particle number fluctuations. In the limit
%of large charge density  the role of the exact charge conservation
%becomes more important than quantum Bose or Fermi effects.

In the present paper we study the particle number fluctuations in
relativistic ideal Bose and Fermi gases for non-zero values of the
net charge density in the GCE and CE.
 The paper is organized as follows. In Sections II and III
we calculate the average values for $N_+$ and $N_-$ and discuss
the Bose condensation in relativistic gases.
These results are not new, and we present them 
in our paper for completeness.
In Section IV we consider the $N_+$ and $N_-$
fluctuations in the GCE and study
the Bose and Fermi effects for particle number densities. 
In Section V the same calculations and study
are repeated within the CE. We compare the GCE and CE results and
summarize our consideration in Section VI.

%An exact charge conservation reduces the values of  $N_+$ and
%$N_-$ fluctuations in the thermodynamic limit. At the non-zero net
%charge $Q=N_+-N_-$ the canonical ensemble predicts a difference
%for the fluctuations of $N_+$ and $N_-$.
% they are also different
%from the fluctuations of all charged particles $N_{ch}=N_++N_-$.
%All these features of the canonical ensemble are in a striking
%difference with those in the grand canonical ensemble.

%Results of Ref.\,\cite{ce-fluc} and the present study demonstrate
%that there are also the canonical ensemble effects for the
%fluctuations. In contrast to the canonical suppression of average
%multiplicities, the canonical effects for the multiplicity
%fluctuations do survive at $V\rightarrow\infty$ and they are even
%most clearly seen in the thermodynamic limit. The changes of the
%scaled variances due to an exact charge conservation of the
%canonical ensemble are not small (about 50 percent effects) and
%they are in general different for positively, negatively and all
%charged particles.

\section{Average particle numbers}
The relativistic ideal  Bose or Fermi gas can be characterized by
the occupation numbers $n_{p}^{+}$  and $n_{p}^{-}$ of the
one-particle states labeled by momenta $p$ for 'positively
charged' particles and 'negatively charged' particles,
respectively. The GCE average values
%and fluctuations
are \cite{lan}:
 \eq{
 \langle n_p^{\pm} \rangle_{g.c.e.}
 ~ = ~\frac {1}
 {\exp \left[\left(\sqrt{p^{2}+m^{2}}~\mp~ \mu\right) / T\right]
 ~-~ \gamma}~, \label{np-aver}
  }
where $m$ is the particle mass, $T$ is the system temperature and
$\mu$ is the chemical potential connected with the conserved
charge $Q$:
\eq{ \label{NQ} Q~\equiv~\langle N_+\rangle_{g.c.e.}~-~ \langle
N_-\rangle_{g.c.e.}~=~\sum_{p}\langle n_p^{+} \rangle_{g.c.e.}~-~
\sum_{p}\langle n_p^{-} \rangle_{g.c.e.}~.
}
 The parameter $\gamma$ in Eq.~(\ref{np-aver}) is equal to $+1$ and $-1$ for Bose and
Fermi statistics, respectively ($\gamma=0$ corresponds to the
Boltzmann approximation).
 Each  level should be
 further specified by the projection of a particle
 spin. Thus, each $p$-level splits into $g=2j+1$
 sub-levels. It will be assumed that the $p$-summation includes all
 these sub-levels too.
 %This does not change the above formulation
 %because of statistical independence of these quantum sub-levels.
 In the thermodynamic limit the system volume $V$ goes to
 infinity, and the degeneracy factor $g$ enters explicitly when one substitutes
the summation over discrete levels by the integration,
$\sum_{p}~...~=~gV(2\pi^{2})^{-1}~\int_{0}^{\infty}p^{2}dp~...~$.
%\nonumber}
 The particle number densities in the GCE are:
 \eq{ \label{ngce}
 \rho_{\pm}
 &\;\equiv~
 \frac{\langle N_{\pm}\rangle_{g.c.e.}}{V}~
 =~\frac{\sum_p\langle n^{\pm}_p\rangle_{g.c.e.}}{V}
 %\; \frac{\sum_{p}\langle
 %n_{p}^{\pm}\rangle_{g.c.e.}}{V}
  \;=\; \frac{g}{2\pi^2}\;
        \int_0^{\infty}\frac{p^2 dp}{\exp \left[\left(\sqrt{p^{2}+m^{2}}~\mp~ \mu\right)
              / T\right] ~-~ \gamma}\nonumber
  \\
 &\;=\; \frac{g\,T^3}{2\pi^2}\;
       \int_0^{\infty}\frac{x^2 dx}{\exp \left[\sqrt{x^{2}+m^{*\,2}}~\mp~ \mu^*
       \right] ~-~ \gamma}\;,
 }
where $m^*\equiv m/T,\;\mu^*\equiv \mu^*/T\;$. To be definite we
consider $\mu^*\ge 0$ in what follows. This corresponds to
non-negative values of the system charge density $\rho_Q\equiv
\rho_+ -\rho_-\ge 0$. Results for $\mu^*\le 0$ can be obtained
from those with $\mu^*\ge 0$ by exchanging of $N_+$ and $N_-$. In
the Boltzmann approximation ($\gamma=0$) one finds:
 \eq{ \label{nBoltz}
 \rho_{\pm}^{Boltz} \;=\;
 \frac{g\,T^3}{2\pi^2}\;m^{*\,2}\,K_2(m^*)\,\exp(\pm\mu^*)
 \;\simeq \;
 \begin{cases}
 g\,T^3\;\exp(\pm\mu^*)/\pi^2\;, \quad &m^*\ll 1
 \vspace{0.2cm}
 \\
 \;g\,T^3\,(m^*/2\pi)^{3/2}\;\exp[-(m^*\mp\mu^*)]\;, \quad &m^*\gg 1
 \end{cases}
 }
where $\;K_2\;$ is a modified Hankel function. For $0\le \mu^*\leq
m^*$ it follows from Eq.~(\ref{ngce}):
 \eq{ \label{nK2}
 \rho_{\pm}\;=\;
  \frac{g\,T^3}{2\pi^2}\;m^{*\,2}\,
        \sum_{n=1}^{\infty} \frac{\gamma^{n-1}}{n}\;K_2(nm^*)\,\exp(\pm
        n\mu^*)~.
}
Note that for $\rho_-^{Fermi}$ the series expansion in
Eq.~(\ref{nK2}) is convergent for all $\mu^*$. The first
term, $n=1$, in Eq.~(\ref{nK2}) corresponds to the Boltzmann
approximation  and others, $n>1$, are the Bose ($\gamma=1$) or
Fermi ($\gamma=-1$) statistics corrections. For any $T$ and $\mu$
these correction terms lead to
%to the
%Bose enhancement,
$\rho_{\pm}^{Bose}>\rho_{\pm}^{Boltz}$ and
%Fermi suppression,
$\rho_{\pm}^{Fermi}<\rho_{\pm}^{Boltz}$ for the particle number
densities. The Bose enhancement and Fermi suppression factors,
\eq{\label{R}
 R_{\pm}^{Bose}~\equiv~\frac{
\rho_{\pm}^{Bose}}{\rho_{\pm}^{Boltz}}~,~~~~
R_{\pm}^{Fermi}~\equiv~\frac{
\rho_{\pm}^{Fermi}}{\rho_{\pm}^{Boltz}}~, }
 for different values of $m^*$ are shown in Fig.~1 as functions of
 $\mu^*$.

\begin{figure}[ht!]
 \hspace{-0.7cm}
 \epsfig{file=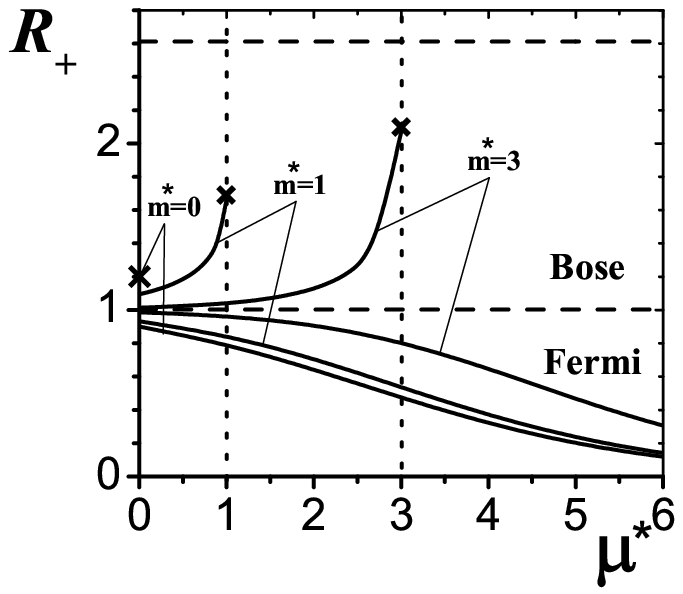,height=6cm,width=8cm}
% \end{figure}
%
% \begin{figure}
\hspace{0.7cm}
 \epsfig{file=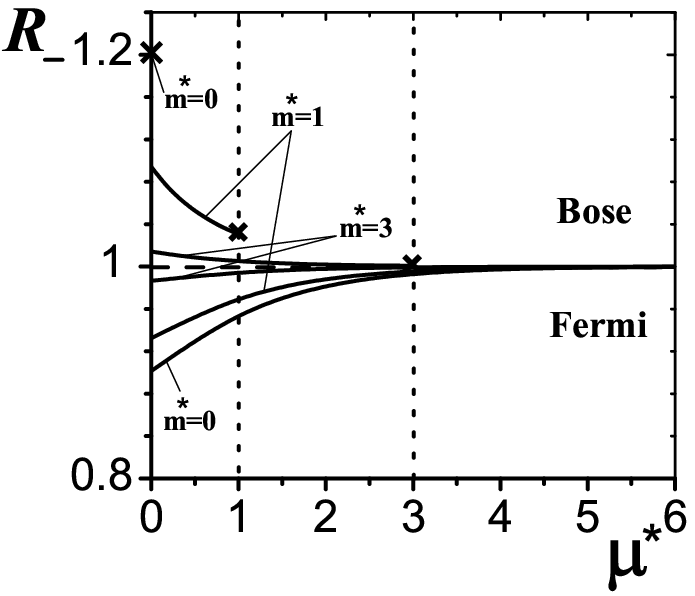,height=6cm,width=8cm}
 \caption{The ratios $R_+$ (left) and $R_-$
 (right) of particle number densities of bosons
($\gamma=1$) and fermions ($\gamma=-1$) to those of classical
particles ($\gamma=0$) are shown as functions of $\mu^*$. Two
upper solid lines show $R_{\pm}^{Bose}$ at $m^*=1,~3$. Three lower
solid lines show $R_{\pm}^{Fermi}$ at $m^*=0,~1,~3$. The vertical
dotted lines $\mu^*=1,~3$ demonstrate the restriction $\mu^*\le
m^*$ in the Bose gas.
% (for  massless case $m^*=0$ only the
%value $\mu^*=0$ is permitted in the Bose gas).
The crosses at the end of the $R_+^{Bose}$ and $R_-^{Bose}$ lines
at $\mu^*=1$ and $\mu^*=3$ correspond to the points of the Bose
condensation. The crosses at $\mu^*=0$ correspond to the limit
$m^*\rightarrow 0$ given by Eq.~(\ref{m0B}) in the Bose gas. The
dashed horizontal line on the left corresponds to the maximum
value  for $R_+^{Bose}$ given by Eq.~(\ref{mueqmp}).
 \label{fig1}
 }
\end{figure}

One finds that the largest  quantum statistics effects  at
$\mu^*=0$ correspond to the massless particles:
\eq{ \label{m0B}
R_{\pm}^{Bose}(\mu^*=0,~ m^*\rightarrow 0)~
%\equiv~\frac{n_{\pm}^{Bose}}{n_{\pm}^{Boltz}}~
&=~
\frac{1}{2}\int_0^{\infty}\frac{x^2dx}{\exp(x)~-~1}~=~\sum_{n=1}^{\infty}\frac{1}{n^3}~
=~\zeta(3)\simeq 1.202~,\\
R_{\pm}^{Fermi}(\mu^*=m^*=0)~
%\frac{n^{\pm~Fermi}}{n^{\pm~Boltz}}~
&=~
\frac{1}{2}\int_0^{\infty}\frac{x^2dx}{\exp(x)~+~1}~=~\sum_{n=1}^{\infty}\frac{(-1)^{n+1}}
{n^3}~ =~\frac{3}{4}\,\zeta(3)\simeq 0.902~, \label{m0F} }
 where
$\zeta(k)$ is a Riemann zeta function (see Appendix A). Note that the values 
of $n_0^+$ and $n_0^-$ contribute to the net charge of the system, but 
for $m=0$ they do not influence the system energy. 
Therefore, the occupation numbers $n_0^+$ and $n_0^-$ become arbitrary, and
the ideal Bose gas of charge particles with $m=0$ has no clear
meaning in the thermodynamic limit.
In what follows the 'massless' Bose gas of charged particles will be
understood as the limit $m^*\rightarrow 0$ at fixed value
of $\mu^*\equiv 0$.

For $m^*\gg 1$ using the asymptotic of the $K_2$ function one
finds:
 \eq{ \label{nminf}
 \rho_{\pm}~
 %=~
 % \frac{g\,T^3}{2\pi^2}\;m^{*\,2}\,
 %       \sum_{n=1}^{\infty} \frac{1}{n}\;K_2(nm^*)\,e^{\pm n\mu^*}
 \simeq ~ \frac{g\,T^3\,m^{*\,3/2}}{(2\pi)^{3/2}}\,
        \sum_{n=1}^{\infty}
        \frac{(\gamma)^{n-1}}{n^{3/2}}~
\exp\left[-n\,(m^*~\mp~\mu^*)\right]
 \;=\; \frac{g\,T^3\,m^{*\,3/2}}{(2\pi)^{3/2}}~
     \gamma ~ Li_{3/2}(\gamma \exp[-(m^*~\mp~\mu^*)])
 }
where
 $
 \sum_{n=1}^{\infty} z^n/n^k
 = Li_k(z)$,
is a polylogarithm function (see Appendix A).
% that is sometimes known as
%Jonqui\`{e}re's function.
For $\;z=1\;$ it equals to the Riemann zeta function,
$\;Li_k(1)=\zeta(k)\;$. The series expansion in Eq.~(\ref{nminf})
for $\rho_+$ converges rapidly  at $\mu^*\ll m^*$. In this case it
is enough to add one term $n=2$ to the Boltzmann approximation
($n=1$) to describe accurately the Bose or Fermi effects. The same
is valid for $\rho_-$ at $m^*\gg 1$ for all $\mu^*$.

%The Eq.(\ref{nminf}) diverges for $\;\rho_+\;$ at $\mu>m$.
The condition $\mu^*\le m^*$ is a general requirement in the Bose
gas. At $\mu^* \rightarrow m^*$ the Bose enhancement factor
$R_+^{Bose}$ reaches its maximum value, and the Bose condensation
of positively charged particles starts. This maximum value of
$R_+^{Bose}$ at $\mu^* = m^*$ increases with $m^*$ and reaches its
upper limit,
\eq{ \label{mueqmp}
%\frac{n_{+}^{Bose}}{n_{+}^{Boltz}}~=~
%
max[R_+^{Bose}(\mu^*,m^*)]~=~R_+^{Bose}(\mu^*=m^*\rightarrow\infty)~=~
Li_{3/2}(1)~=~\zeta(3/2)~\simeq 2.612~, }
at $m^*\rightarrow \infty$ (see Fig.~1, left).  For negatively
charged particles the Bose enhancement factor reaches its minimal
value at $\mu^*=m^*$,
\eq{
\label{mueqmm}
%\frac{n_{-}^{Bose}}{n_{-}^{Boltz}}
R_-^{Bose}(\mu^*=m^*)~\simeq~1~+~\frac{K_2(2m^*)}{2K_2(m^*)}~\exp\left(-m^*\right),
}
and this value goes to 1 (i.e. to its classical Boltzmann limit)
from above at $m^*\rightarrow \infty$ (see Fig.~1, right).

 The requirement $\mu^*\le m^*$ is absent  for the Fermi
gas and for $\mu \gg m$ one finds (see Eq.~(\ref{A6}) in Appendix
B):
 \eq{ \label{nplusF}
 \rho_{+}^{Fermi}
% &\;=\; \frac{g\,T^3}{2\pi^2}\,
%        \int_0^{\infty}\frac{x^2 dx}{\exp \left[\sqrt{x^{2}+m^{*\,2}}~\mp~ \mu^*
%        \right] ~+~ 1}\nonumber
% \\
% \simeq\; \frac{g\,T^3}{2\pi^2}\,
%  \left[\; \frac{1}{3}\left(\mu^{*\,2}-m^{*\,2}\right)^{3/2}
%        +  \frac{\pi^2}{6}\;\frac{2\mu^{*\,2}-m^{*\,2}}{\sqrt{\mu^{*\,2}-m^{*\,2}}}\;
%        \right]~
~  \simeq~\frac{g\,T^3}{2\pi^2}\,
  \left[\;\frac{1}{3}\;\mu^{*\,3} + \left(\frac{\pi^2}{3}
         -\frac{m^{*\,2}}{2}\right)\mu^*\right]
         \, ,
 }
while the density for negatively charged particles can be
approximated in this limit  with the first two terms from the
right-hand side of Eq.~(\ref{nK2}): $n=1$ corresponds to the
Boltzmann approximation $\rho_{-}^{Boltz}$ ~(\ref{nBoltz}), and
$n=2$ gives a small (negative) Fermi correction.  Thus, one finds
that $R_+^{Fermi}$ goes to zero (see Fig.~1, left) and
$R_-^{Fermi}$ goes to 1 from below (see Fig.~1, right) at
$\mu^*\rightarrow\infty$:
 \eq{
 \label{RFpm}
R_+^{Fermi}~\simeq~\frac{\mu^{*3}\,\exp\left(-\mu^*\right)}{3~m^{*2}K_2(m^*)}
 ~\rightarrow~0~,~~~~
R_-^{Fermi}~\simeq~1~-~\frac{K_2(2m^*)}{2K_2(m^*)}~\exp\left(-\mu^*\right)
\rightarrow~1~. }

%\newpage

%
%%%%%%%%%%%%%%%%%%%%%%%%%%%%%%%%%%%%%%%%%%%%%%%%%%%%%%%%%%%%%%%%%%%%%%%%%%%%%%%%%%%%%%%%
%
%%%%%%%%%%%%%%%%%%%%%%%%%%%%%%%%%%%%%%%%%%%%%%%%%%%%%%%%%%%%%%%%%%%%%%%%%%%%%%%%%%%%%%%%
%\newpage
\section{Bose condensation}
In a standard non-relativistic picture of the Bose condensation
the particle number $N$ is a conserved quantity. If the system
temperature decreases at fixed particle number density $\rho=N/V$,
the system chemical potential increases and reaches its maximal
value at $T=T_C$.  Bose condensation starts and a macroscopic part
of the system particles -- known as the Bose condensate --
occupies the lowest momentum state at $T<T_C$. In a relativistic
picture the conserved quantity is system charge $Q=N_+-N_-$. From
Eq.~(\ref{nK2}) one finds for the dimensionless charge density:
 \eq{ \widetilde{\rho}_Q~\equiv~
 %\widetilde{\rho}_+~-~\widetilde{\rho}_-~=~
\frac{\rho_Q}{g\,m^{3}}~\equiv
 ~\frac{1}{g\,m^{3}}~\left(\rho_+^{Bose}~-~\rho_{-}^{Bose}\right) \;=\;
 \frac{1}{\pi^2}~
 \sum_{n=1}^{\infty}\frac{1}{n\,m^*}~\;K_2\left(n\,m^*\right)~\sinh(n\,\mu^*)~.
 \label{rhoQ}
 }
%where
%$\widetilde{T}_C\equiv T_C/m$,~
% $\widetilde{\rho}_{\pm}\equiv\rho_{\pm}/(gm^3)$.

Bose condensation starts at the point $\;T=T_C\;$ when
$\;\mu=\mu^{max}=m\;$. At this point Eq.~(\ref{rhoQ}) is reduced
to:
 \eq{ \widetilde{\rho}_Q~
 %\equiv~\widetilde{\rho}_+~-~\widetilde{\rho}_-~
 =~
 %\frac{\rho_{\pm}}{g\,m^3} \;=\;
 \frac{\widetilde{T}_C}{\pi^2}~
 \sum_{n=1}^{\infty}\frac{1}{n}\;
 K_2\left(n/\widetilde{T}_C\right)~\sinh(n/\widetilde{T}_C)~,
 \label{TC}
 }
 where $\widetilde{T}_C\equiv T_C/m$.
% $\widetilde{\rho}_{\pm}\equiv\rho_{\pm}/(gm^3)$.
Equation (\ref{TC}) can  be used to write the Bose condensation
temperature $T_C$ as the function of the conserved charge density
$\rho_Q$. The line of  Bose condensation
$\widetilde{T}_C=\widetilde{T}_C(\widetilde{\rho}_Q)$ given by
Eq.~(\ref{TC}) is shown in Fig.~2 (see also Ref.~\cite{TC} and
references therein).
\begin{figure}[ht!]
\epsfig{file=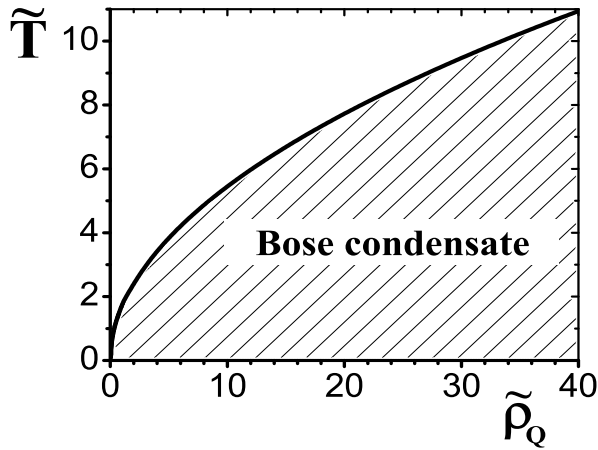,height=6cm,width=8cm}
 %\hspace{0.7cm}
 %\epsfig{file=trholog2.eps,height=5cm,width=8cm}
 \caption{The phase diagram of the relativistic ideal Bose gas.
 The solid line  shows Bose
condensation  temperature as a function of the conserved charge
density. It is given by Eq.(\ref{TC}) where both quantities are
expressed in dimensionless form: $\widetilde{T}_C\equiv
T_C/m,~\widetilde{\rho}_Q\equiv \rho_Q/(gm^{3})$. At
$\widetilde{T}_C\ll 1$ the line
$\widetilde{T}_C(\widetilde{\rho}_Q)$ is given by non-relativistic
approximation (\ref{TCs}), while at $\widetilde{T}_C\gg 1$ it is
described by the ultra-relativistic relation (\ref{TCl}). The
points $(\widetilde{\rho}_Q, \widetilde{T})$ under the solid line
correspond to the states of the system with non-zero values of the
Bose condensate.
%The dashed lines in the right
%panel correspond to the non-relativistic approximation
%(\ref{TC23}) and ultra-relativistic one (\ref{TC12}).
 \label{fig2}}
\end{figure}
%
%
%
% \eq{
% \frac{\rho_{Q}}{g\,m^3} \;\equiv\; \frac{\rho_+-\rho_-}{g\,m^3}
% \;=\; \frac{a^3}{\pi^2} \sum_{n=1}^{\infty}\frac{1}{n}\;
%       K_2\left(n/a\right)\,\sinh(n/a)
% }
%

At $\widetilde{T}_C\ll 1$ it follows from Eq.~(\ref{TC}):
 \eq{
 \widetilde{\rho}_{Q} \;\simeq\; \left(\frac{\widetilde{T}_C}{2\pi}\right)^{3/2}
 \sum_{n=1}^{\infty}\frac{1}{n^{3/2}} \;=\;
 \left(\frac{\widetilde{T}_C}{2\pi}\right)^{3/2}\,\zeta(3/2)~.
 \label{TCs}
}
This corresponds to a non-relativistic limit.
% of the Bose
%condensation.
The density of negatively charged particles at $T\simeq T_C$
behaves as $\rho_- \propto \exp(-2/\widetilde{T}_C)$ and can be
neglected in a comparison with $\rho_+ \propto
(\widetilde{T}_C)^{3/2}$. Under these conditions the charge number
conservation becomes equivalent just to the (positively charged)
particle number conservation. One, therefore, recovers from
Eq.~(\ref{TCs}) the familiar relation, $\widetilde{T}_C \simeq
3.313 \widetilde{\rho}_Q^{~2/3}$, between the Bose condensation
temperature and particle number density known in the
non-relativistic statistical mechanics \cite{lan}. At
$\widetilde{T}_C\gg 1$ from Eq.~(\ref{TC}) one finds:
 \eq{
 \widetilde{\rho}_{Q}\;\simeq\;
 \frac{2\,
 \widetilde{T}_C^2}{\pi^2}\sum_{n=1}^{\infty}\frac{1}{n^2}
 %\;=\; \frac{2\,\widetilde{T}_C^2}{\pi^2}\,\zeta(2)
 \;=\; \frac{\widetilde{T}_C^2}{3}~.
 \label{TCl}
}
This corresponds to  the ultra-relativistic limit  and
%, instead of
%Eq.~(\ref{TC23}),
leads to a new relation, $\widetilde{T}_C \simeq 1.732
\widetilde{\rho}_Q^{~1/2}$, between the Bose condensation
temperature and charge number density.

 For fixed value of the conserved charge density $\rho_Q$
the chemical potential is constant $\mu^*=m^*$ at $T\le T_C$. The
positively charged particles have to condensate at the lowest
quantum level to preserve a constant value of the positive charge
density in the system. Therefore, one finds at $T\le T_C$:
\eq{
\rho_Q~=~\rho_+^{Bose}(T,\mu^*=m^*)~-~\rho_-^{Bose}(T,\mu^*=m^*)~+
~\rho_{+}^{cond}(T)~, \label{cond}
}
where the first two terms in Eq.~(\ref{cond}) are given by
Eq.~(\ref{ngce}),
% with $\gamma=1$,
and $\rho_{+}^{cond}$ is the density of positively charged
particles at the lowest quantum level (Bose condensate). The
behavior of $\rho_+^{Bose}$, $\rho_-^{Bose}$ and $\rho_+^{cond}$
above and below the Bose condensation temperature are shown in
Fig.~3. Note that $\rho_+^{Bose}$ and $\rho_-^{Bose}$ are
calculated by Eq.~(\ref{ngce}) with $\mu^*\le m^*$ at $T\ge T_C$
(the value of $\mu^*$ is defined by the equation
$\rho_Q=\rho_+^{Bose}-\rho_-^{Bose}$) and with $\mu^*=m^*$ at
$T<T_C$ ($\rho_+^{cond}$ is given by Eq.~(\ref{cond}) at $T<T_C$
and it equals to zero at $T>T_C$). The Bose condensation is the
3$^{rd}$ order phase transition with a maximum of specific heat at
$T=T_C$.
\begin{figure}[ht!]
\hspace{-0.7cm}
 \epsfig{file=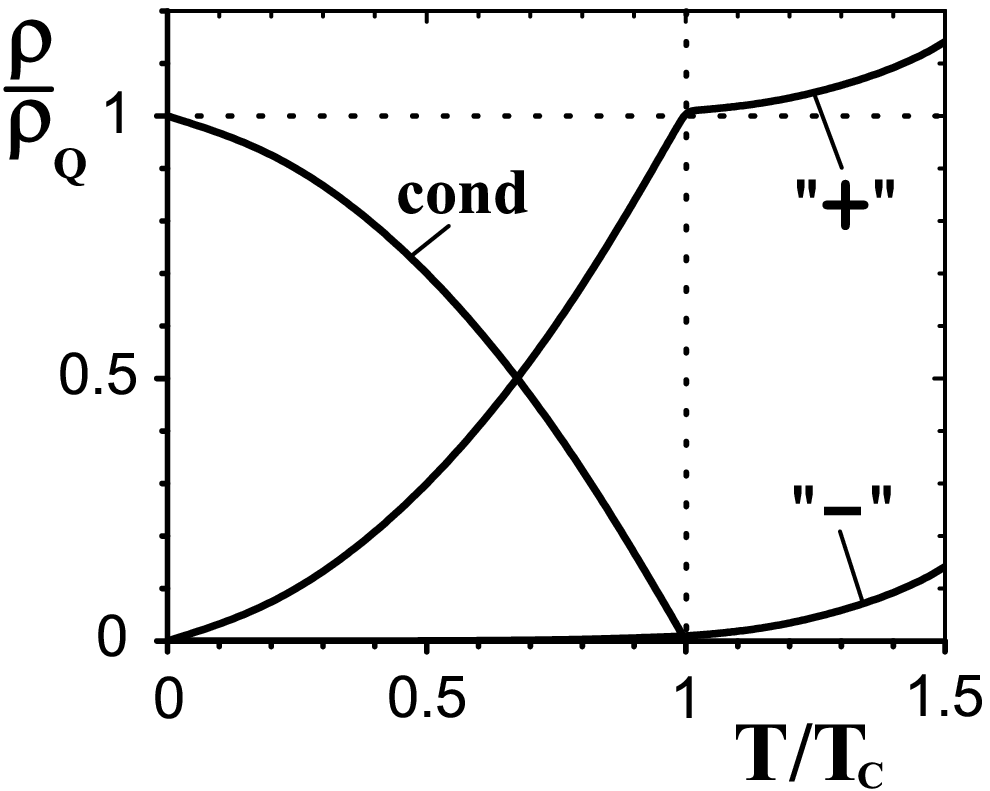,height=6cm,width=8cm}
%\end{figure}
%
\hspace{0.7cm}
%\begin{figure}[h!]
%\hspace{-0.7cm}
 \epsfig{file=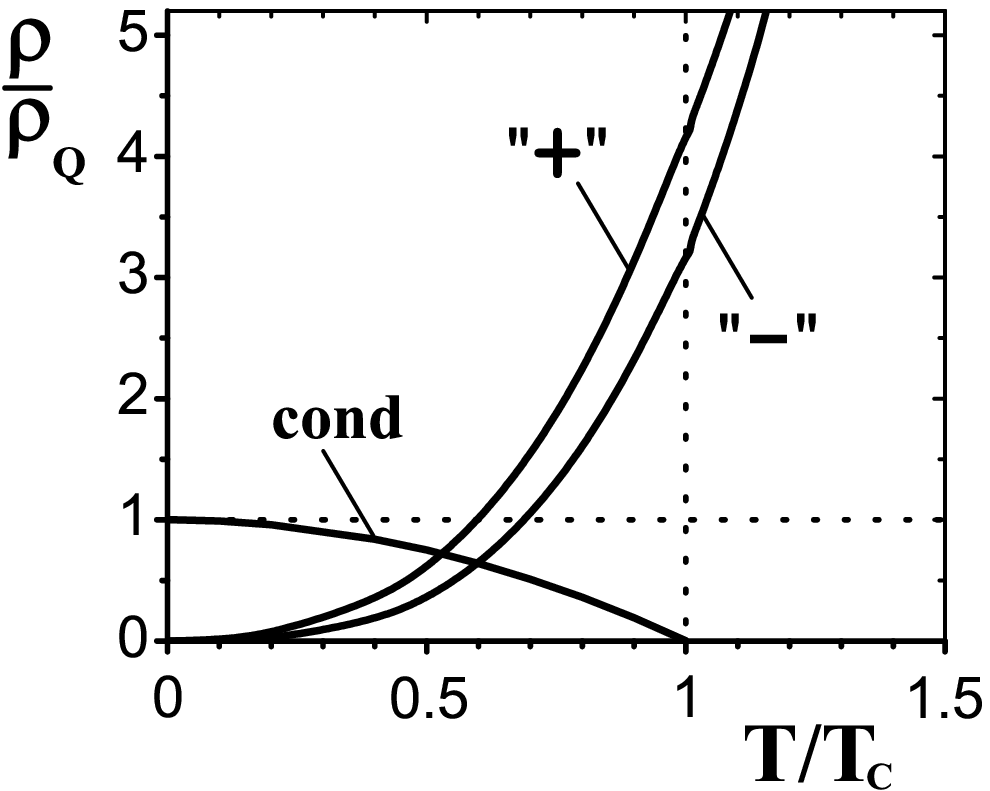,height=6cm,width=8cm}
\caption{The  solid lines show the ratios $\rho_+^{Bose}/\rho_Q$,
$\rho_-^{Bose}/\rho_Q$, and $\rho_+^{cond}/\rho_Q$ as functions on
$T/T_C$ at fixed values of $\rho_Q$ for the relativistic ideal
Bose gas. The left and right pictures correspond to the values
$\widetilde{T}_C=0.5,~ \widetilde{\rho}_Q\simeq 0.06$, and
$\widetilde{T}_C=10,~\widetilde{\rho}_Q\simeq 33.3$, respectively.
Both these  $(\widetilde{\rho}_Q, \widetilde{T}_C)$
 points belong to the Bose condensation line in
Fig.~2 --  the first point lies close to the lower left corner and
 the second point lies close to the upper right corner in
Fig.~2. The system presented in the left picture can be treated
within non-relativistic approximation. In this case, $\rho_Q\simeq
\rho_+^{Bose}$  and $\rho_-^{Bose}/\rho_+^{Bose}\ll 1$, so that
negatively charged particles can be neglected at $T<T_C$ and
charge conservation becomes equivalent to particle number
conservation. The system presented in the right picture
demonstrates the Bose condensation in the ultra-relativistic case:
both $\rho+^{Bose}$ and $\rho_-^{Bose}$ are essentially larger
than conserved charge density $\rho_Q$ in the vicinity of the Bose
condensation temperature $T=T_C$.} \label{fig3}
\end{figure}

%
%%%%%%%%%%%%%%%%%%%%%%%%%%%%%%%%%%%%%%%%%%%%%%%%%%%%%%%%%%%%%%%%%%%%%%%%%%%%%%%%%%%%%%%%
%
%%%%%%%%%%%%%%%%%%%%%%%%%%%%%%%%%%%%%%%%%%%%%%%%%%%%%%%%%%%%%%%%%%%%%%%%%%%%%%%%%%%%%%%%
%\newpage
\section{Particle number fluctuations in the GCE}
The GCE fluctuations of the single-mode occupation numbers are
equal to \cite{lan}:
 \eq{
\langle\Delta n_p^{\pm 2}\rangle_{g.c.e.}  \equiv \langle
\left(n_p^{\pm}-\langle
n_p^{\pm}\rangle_{g.c.e.}\right)^2\rangle_{g.c.e.}~=~ \langle
n_{p}^{\pm~2}\rangle_{g.c.e.}~-~ \langle
n_{p}^{\pm}\rangle^{2}_{g.c.e.}~=~
%%~T\frac{\partial n_p}{\partial
%%  \mu}=
\langle n_p^{\pm} \rangle_{g.c.e.} \left(1 + \gamma \langle
n_p^{\pm} \rangle_{g.c.e.}\right) ~ \equiv~
  v^{\pm~ 2}_p~.\label{np-fluc}
  }
 The fluctuations
  of the macroscopic observables can be written
  in terms of the microscopic correlator $\langle \Delta n_p^\alpha \Delta
  n_k^\beta \rangle_{g.c.e.}$, where $\alpha,\beta$  are $+$
  and/or
  $-$, which has a simple form,
  \eq{
  \langle \Delta n_p^\alpha \Delta n_k^\beta \rangle_{g.c.e.}~ =~
  v_p^{\alpha~2}~
  \delta_{p\,k}~\delta_{\alpha \beta}~,  \label{correlator1}
  }
   due to the statistical independence of different
  quantum levels and different charge states in the GCE.
 The variances of the
 total number of positively and/or negatively charged  particles
 are equal to:
 \eq{
 \langle \Delta N_{\pm}^{2}\rangle_{g.c.e.}\,  \equiv\,
 \langle N_{\pm}^{2} \rangle_{g.c.e.} - \langle N_{\pm}
 \rangle^2_{g.c.e.}\,=\, \sum_{p,k} \langle n_p^{\pm} n_k^{\pm}
 \rangle_{g.c.e.} -
  \langle n_p^{\pm} \rangle_{g.c.e.} \langle n_k^{\pm} \rangle_{g.c.e.}
 = \sum_{p,k} \langle \Delta n_{p}^{\pm} \Delta_k^{\pm}
 \rangle_{g.c.e.}\, = \, \sum_p v_p^{\pm~ 2}~.
\label{deltaN2gce}
}

The scaled variance $\omega^{\pm}_{g.c.e.}$
% in the\mapsto
%thermodynamical limit $V\rightarrow\infty$
reads:
 \eq{\label{omegagce1}
 \omega^{\pm}_{g.c.e.} ~&\equiv~\frac{\langle N_{\pm}^2
 \rangle_{g.c.e.}~-
 ~\langle N_{\pm}\rangle^{2}_{g.c.e.}}
 {\langle N_{\pm} \rangle_{g.c.e.}}~=~ \frac{\sum_{p,k} \langle
 \Delta n_{p}^{\pm} \Delta n_{k}^{\pm} \rangle_{g.c.e.}}
 {\sum_p \langle n^{\pm}_p\rangle_{g.c.e.}}~ = ~
 \frac{\sum_{p}v_p^{\pm ~2}}{V~\rho_{\pm}}
 %\sum_{p}\langle n_p^\alpha\rangle_{g.c.e.}}
 %~\simeq~
%\frac{\int_{0}^{\infty}p^{2}dp~v_p^{\alpha
%2}}{\int_{0}^{\infty}p^{2}dp~\langle n_p^\alpha\rangle_{g.c.e.}}
\nonumber \\
&= ~ 1\;+\;\gamma\;
%       \frac{\int_{0}^{\infty}p^{2}dp~\langle n_p^\alpha\rangle^2_{g.c.e.}}
%            {n^{\alpha}}~=~1~+~,
    \int_0^{\infty}\frac{x^2 dx}{\left[\;\exp \left(\sqrt{x^{2}+m^{*\,2}}~\mp~ \mu^*
       \right) ~-~ \gamma\;\right]^2}~\times~\left[
       \int_0^{\infty}\frac{x^2 dx}{\;\exp
       \left(\sqrt{x^{2}+m^{*\,2}}~\mp
       ~ \mu^*
       \right) ~-~ \gamma}\right]^{-1}~,
 }
where the thermodynamic limit is assumed, and the $p$-summation is
substituted by the integration similar to Eq.~(\ref{ngce}).
%%%%%%%%%%%%%%%%%%%%%%%%%%%%%%%%%%%%%%%%%%%%%%%%%%%%%%%%%%%%%%%%%%%%%%%%%%%%%%%%%%%%%%%%%%%%%
%
The scaled variances $\omega_{g.c.e.}^{\pm Bose}$ and
$\omega_{g.c.e.}^{\pm Fermi}$  for different values of $m^*$ are
shown in Fig.~4 as functions of $\mu^*$.

%\newpage
\begin{figure}[ht!]
 \hspace{-0.7cm}
 \epsfig{file=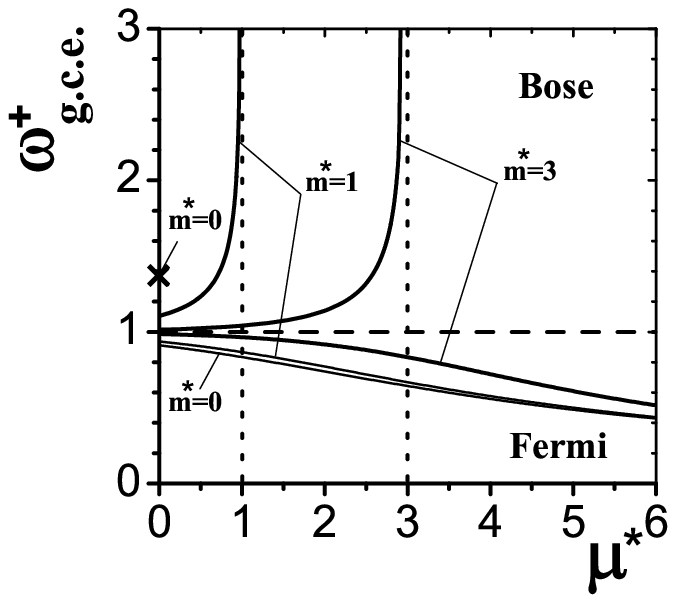,height=6cm,width=8cm}
% \end{figure}
%
% \begin{figure}
%\epsfig{file=fig4b.eps,height=10cm,width=16cm}
\hspace{0.7cm}
 \epsfig{file=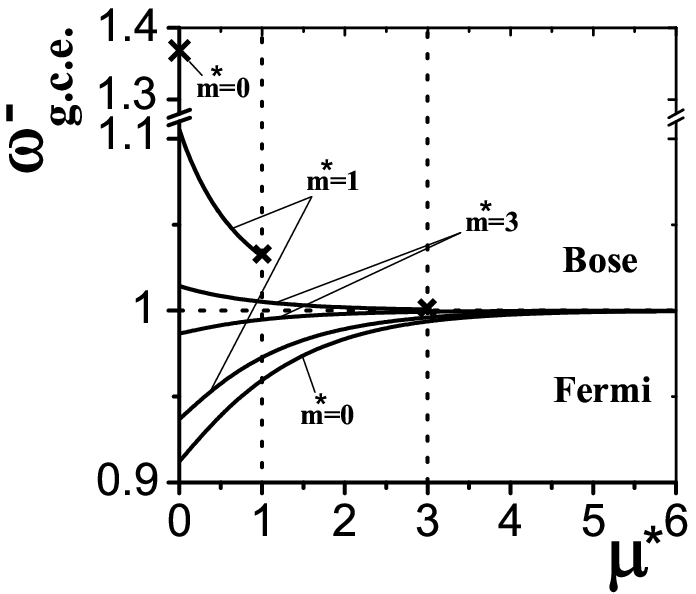,height=6cm,width=8cm}
 \caption{The scaled
variances $\omega_{g.c.e.}^+$ (left) and $\omega_{g.c.e.}^-$
 (right) given by Eq.~(\ref{omegagce1}) for bosons
($\gamma=1$) and fermions ($\gamma=-1$) are shown as functions of
$\mu^*$.  The two upper solid lines present $\omega_{g.c.e.}^{\pm
Bose}$ for $m^*=1,~3$.  The three lower solid lines present
$\omega_{g.c.e.}^{\pm Fermi}$ for $m^*=0,~1,~3$.  The vertical
dotted lines $\mu^*=1,~3$ demonstrate the restriction $\mu^*\le
m^*$ in the Bose gas.
% ( for the massless case $m^*=0$ only the value
%$\mu^*=0$ is permitted in the Bose gas).
The crosses at the end of the lines  for $\omega_{g.c.e.}^{-Bose}$
at $\mu^*=1$ and $\mu^*=3$ correspond to the points of the Bose
condensation, $\omega_{g.c.e.}^{+Bose}$ diverges at these points.
The crosses at $\mu^*=0$ correspond to the limit $m^*\rightarrow
0$ given by Eq.~(\ref{m0Bfluc}) in the Bose gas. \label{fig4}}
\end{figure}
It follows from Eq.~(\ref{omegagce1}) for $\gamma=0$,
\eq{\label{omegagce}
\omega^{+~Boltz}_{g.c.e.}~=~ \omega^{-~Boltz}_{g.c.e.}~=~1~, }
i.e. the scaled variances for Boltzmann statistics in the GCE are
independent of the chemical potential $\mu^*$ and equal to 1 for
both the positively and negatively charged particles. The
Eq.~(\ref{omegagce1}) leads to the Bose enhancement,
$\omega_{g.c.e.}^{\alpha ~Bose}>1$, and the Fermi suppression
$\omega_{g.c.e.}^{\alpha~Fermi}<1$, of the particle number
fluctuations.

 At $\mu^*=0$ the largest Bose and Fermi effects
correspond to the massless particles (see Fig.~4):
\eq{
\label{m0Bfluc} \omega^{\pm~Bose}_{g.c.e}(\mu^*=0,~ m^*\rightarrow
   0)~&=~1~+~
\int_0^{\infty}\frac{x^2dx}{\left[\;\exp(x)~-~1\;\right]^2}
 ~\times~\left[\int_0^{\infty}\frac{x^2dx}
{\exp(x)~-~1}\right]^{-1}~ =~\frac{\zeta(2)}{\zeta(3)}
~\simeq ~ 1.368~,\\
\omega^{\pm~Fermi}_{g.c.e}(\mu^*=m^*=0)~&=~1~-~
\int_0^{\infty}\frac{x^2dx}{\left[\;\exp(x)~+~1\;\right]^2}
 ~\times~\left[\int_0^{\infty}\frac{x^2dx}
{\exp(x)~+~1}\right]^{-1}~
=~\frac{2}{3}\;\frac{\zeta(2)}{\zeta(3)} ~\simeq ~ 0.912~,
\label{m0Ffluc} }

Using Eqs.~(\ref{a5}-\ref{a6}) one finds from
Eq.~(\ref{omegagce1}) at $\mu^*\le m^*$:
\eq{
\label{n2}
% \int_0^{\infty}\frac{x^2 dx}{\left[\;\exp \left(\sqrt{x^{2}+m^{*\,2}}~-\alpha~ \mu^*
%       \right) ~-~ \gamma\;\right]^2}
 \omega_{g.c.e.}^{\pm}~=~1~+~\gamma~\sum_{n=1}^{\infty}\frac{\gamma^{n-1}~n}{n+1}
        \;K_2\left[(n+1)\,m^*\right]\,\exp\left[\pm~ (n+1)\,\mu^*\right]~\times~
\left[\sum_{n=1}^{\infty} \frac{\gamma^{n-1}}{n}
        \;K_2\left(n m^*\right)\,\exp\left(\pm~
        n\mu^*\right)\right]^{-1}~.
}
Note that  Eq.~(\ref{n2}) is valid  for
$\omega_{g.c.e.}^{-~Fermi}$ for all values of  $\mu^*>0$.  At
$m^*\gg 1$ one finds from Eq.~(\ref{n2}):
\eq{\label{omegagce2}
\omega^{\pm}_{g.c.e}~&\simeq~1~+~\gamma~
 \sum_{n=1}^{\infty}\frac{\gamma^{n-1}~n}{(n+1)^{3/2}}~\exp[-(n+1)(m^* ~\mp~\mu^*)]
 ~\times~
 \left[
 \sum_{n=1}^{\infty}\frac{\gamma^{n-1}}{n^{3/2}}~\exp[-n(m^*~\mp~\mu^*)]
 \right]^{-1}~.
 }
The series expansions in Eq.~(\ref{omegagce2}) converge rapidly
for $\mu^*\ll m^*\rightarrow\infty$. In this case  the term with
$n=1$ is sufficient to describe small Bose or Fermi effects:
\eq{
\label{omegagce3}
\omega^{\pm}_{g.c.e}~\simeq~1~+~\gamma~2^{-3/2}~\exp[-(m^*~\mp~\mu^*)]~.
 }
 The same is valid for negatively charged particles at
 $\mu^*\rightarrow\infty$:
 \eq{\label{omegagce4}
 \omega^{-}_{g.c.e}~\simeq~1~+~\gamma~\frac{K_2(2m^*)}{2K_2(m^*)}~
 \exp[-~\mu^*)]~. }
The first terms in Eqs.~(\ref{omegagce3}-\ref{omegagce4})
correspond to the Boltzmann scaled variances (\ref{omegagce}).
Therefore, for both positively and negatively charged particles,
the Bose and Fermi corrections approach to zero as $\gamma
\exp(-m^*)$ at $\mu^*\ll m^*\rightarrow\infty$. For negatively
charged particles, these corrections also tend to zero as $\gamma
\exp(-\mu^*)$ at $\mu^*\rightarrow\infty$.

The condition $\mu^*\le m^*$ is a general requirement in the Bose
gas. At $\mu^* \rightarrow m^*$ one the scaled variance
$\omega^{+~Bose}_{g.c.e}$ diverges (see Fig.~4, left).
% and the Bose condensation
%starts.
%From Eq.~(\ref{omegagce1}) one finds
%$\omega^{+Bose}_{g.c.e}\propto(m^*-\mu^*)^{-1/2}\rightarrow\infty$
%at $\mu^* \rightarrow m^*$.
%It can be easily seen from Eq.~(\ref{n2})
%that the scaled variance $\omega^+_{g.c.e.}$ for Bose statistics
%$(m-\mu)^{-1/2}$ at $\mu^*\rightarrow m^*$.
This divergence comes from the contributions of the low momentum
modes. Introducing  a dimensionless parameter $\delta$ satisfying
the conditions $m^*-\mu^*\ll \delta\ll m^*$ one finds:
\eq {
\label{n2bose}
 \int_0^{\delta}\frac{x^2 dx}{\left[\;\exp \left(\sqrt{x^{2}+m^{*\,2}}~-~ \mu^*
       \right) ~-~ 1\; \right]^2}~\simeq~
       %\frac{1}{m^*~-~\mu^*}~
        \int_0^{\delta}\frac{x^2
        dx}{(m^*-\mu^*~+~x^2/2m^*)^2}~\simeq~
\pi~2^{-1/2}~m^{*~3/2}~(m^*~-~\mu^*)^{-1/2}~. }
Therefore,  it follows, $ \omega^{+~Bose}_{g.c.e} \propto
(m^*-\mu^*)^{-1/2}\rightarrow \infty$,  as $\mu^*\rightarrow m^*$.
On the other hand, the scaled variance for negative Bose particles
decreases with $\mu^*$ and reaches its minimum at $\mu^*=m^*$.
When $\mu^*=m^*\rightarrow\infty$ one finds from
Eq.~(\ref{omegagce3}),
\eq{\label{Bmgce}
\omega^{-~Bose}_{g.c.e.}~\simeq
~1~+~2^{-3/2}~\exp\left(-2m^*\right)~, }
 so that $\omega^{-~Bose}_{g.c.e.}$ approaches to 1  from above
 as
 $\mu^*=m^*\rightarrow\infty$ (see Fig.~4, right).
%
%\subsection{\large\bf $\large\bf \mu^*\gg m^*,\; \gamma=-1$, Fermi
%statistic}
%

 The requirement $\mu^*\le m^*$ is absent in the Fermi gas, and
for $\mu^*\rightarrow\infty$ one finds strong Fermi suppression
effects (see Fig.~4, left) for positively charged particles (see
Eq.~(\ref{A14}) in Appendix B):
 \eq{\label{Fpgce}
 \omega^{+~Fermi}_{g.c.e} ~\simeq~
 \frac{3}{\mu^*}\;.
 }
The scaled variance for negatively charged Fermi particles
 increases with $\mu^*$, and from
Eq.~(\ref{omegagce4}),
\eq{\label{Fmgce}
 \omega^{-~Fermi}_{g.c.e}~\simeq~1~-~\frac{K_2(2m^*)}{2K_2(m^*)}~
\exp\left(-~\mu^*\right)~, }
 so that $\omega^{-~Fermi}_{g.c.e}$  approaches to 1 from below
 at $\mu^*\rightarrow\infty$  (see Fig.~4, right).

%

%

%\newpage
\section{Particle number fluctuations in the CE}

In the GCE all possible sets of the occupation numbers
$\{n_p^+,n_p^-\}$ contribute to the partition function. Only the
average value of the conserved charge   $Q=   \sum_p (n_p^+ -
n_p^-)$  is fixed, $\langle Q\rangle_{g.c.e.}=Q$, in the GCE, and
$\langle Q\rangle_{g.c.e.}$    is controlled   by the chemical
potential $\mu^*$.  In the CE an exact charge conservation is
imposed. This can be formulated as a restriction on permitted sets
of the occupation numbers $\{n_p^+,n_p^-\}$: only those satisfying
the relation,
\eq{\label{deltaQ}
   \Delta Q~=~\sum_{p} \left( ~\Delta n_p^{+}~-~\Delta n_p^-~\right)~=~0~
   ,}
contribute to the CE partition function. One proves that this
restriction does not change the average quantities in the
thermodynamic limit, if  the average charge in the GCE, $\langle
Q\rangle_{g.c.e.}$,~  equals the charge $Q$ of the CE (of course,
$T$ and $V$ values are assumed to be the same in the GCE and CE).
In particular,
 \eq{ \langle N_+\rangle_{c.e.}~=~\langle N_+\rangle_{g.c.e.}~,~~~~
\langle N_-\rangle_{c.e.}~=~\langle N_-\rangle_{g.c.e.}~.
 }
 This is what the thermodynamical equivalence of the CE and
GCE  means as $V\rightarrow \infty$. This statistical equivalence
does not apply, however, for the fluctuations, measured in terms
of $\omega^+$ and $\omega^-$. The formula (\ref{correlator1}) for
the microscopic correlator is modified if we impose the
restriction of an exact charge  conservation in a form of
Eq.~(\ref{deltaQ}). One finds (see the details in
Ref.~\cite{ce2-fluc}) the CE correlator:
 \eq{
 \label{corr-ch}
   \langle \Delta n^{\alpha}_p \Delta n^{\beta}_k \rangle_{c.e.} ~ =~
   \delta_{p\,k}~ \delta_{\alpha \beta}~ v_p^{\alpha~ 2}~ -~
   \frac{v^{\alpha~ 2}_p q^{\alpha}~v^{\beta ~2}_k  q^{\beta}}
   {\sum_{p,\alpha} v^{\alpha ~2}_p
   %q^{\alpha 2}
   }~.
   }
  By means of Eq.\,(\ref{corr-ch}) we obtain:
   \eq{
   \omega^{\alpha}_{c.e.}~\equiv~\frac{
    \langle N_{\alpha}^2 \rangle_{c.e.} ~-~ \langle N_{\alpha} \rangle^2_{c.e.}}
    {\langle N_{\alpha} \rangle_{c.e.}} ~=~
\frac{\sum_{p,k} \langle
 \Delta n_{p}^\alpha \Delta n_{k}^\alpha \rangle_{c.e.}}
 {\sum_p \langle n^\alpha_p\rangle_{c.e.}}~=~
    \frac{
    \sum_p v_{p}^{\alpha ~2}}{V~\rho_{\alpha}}\left(1~-~
    %\sum_{p}\langle n_{p}^{\alpha}\rangle_{g.c.e.}}
        \frac{\sum_p v_{p}^{\alpha~ 2}}
        %\sum_{p}\langle n_{p}^{\alpha}\rangle_{g.c.e.}
    {\sum_{p} v_{p}^{+~2}~+~\sum_p v_p^{-~2}}
    %q^{\alpha 2}}
    \right) ~. \label{omega-alpha-ce}
   }
Comparing Eq.\,(\ref{corr-ch}) and Eq.\,(\ref{correlator1}) one
notices the changes of the microscopic correlator due to an exact
charge conservation. Namely, in the CE the fluctuations of each
mode are reduced, and the (anticorrelations) correlations  between
different modes $p\neq k$ with the (same) different  charge states
$\alpha, \beta$ appear. These two changes of the microscopic
correlator result in a suppression of the CE scaled variances
$\omega^{\alpha}_{c.e.}$  in comparison with the GCE ones
$\omega^{\alpha}_{g.c.e.}$ (compare Eq.\,(\ref{omega-alpha-ce})
and Eq.\,(\ref{omegagce1})), i.e. the fluctuations of both $N_+$
and $N_-$ are always smaller in the CE than those in the GCE. A
nice feature of Eq.~(\ref{omega-alpha-ce}) is the fact that
particle number fluctuations in the CE, being different from those
in the GCE, are presented in terms of $\rho_{\pm}$ and
$v_p^{\pm~2}$ given by Eqs.~(\ref{ngce}) and (\ref{np-fluc}),
respectively, both quantities calculated in terms of $\langle
n_p^{\pm}\rangle_{g.c.e.}$ within the GCE.

The Eq.\,(\ref{np-fluc}) leads to $v_{p}^{\alpha ~2}=\langle
n_{p}^{\alpha}\rangle_{g.c.e.}$ in the Boltzmann approximation, so
that $\sum_p v_{p}^{\alpha~ 2}=V\rho_{\alpha}^{Boltz}$, and from
Eq.\,(\ref{omega-alpha-ce}) one finds (see dashed lines in Figs.~5
and 6):
\eq{\omega^{\pm~Boltz}_{c.e.}~=~1~-~\frac{\exp(\pm~\mu^*)}
{\exp(\mu^*)~+~\exp(-\mu^*)}~=~ \frac{1}{2}\left[~1~\mp
~\tanh(\mu^*)\right]~.\label{omega-alpha1}
}
The Eq.~(\ref{omega-alpha1}) demonstrates the CE suppression
effects for particle number fluctuations within the Bolzmann
approximation, e.g., the scaled variances
$\omega_{c.e.}^{+~Boltz}$ and $\omega_{c.e.}^{-~Boltz}$ in the CE
at zero net charge density are two times smaller,
$\omega_{c.e.}^{+~Boltz}=\omega_{c.e.}^{-~Boltz}=0.5$, than those
in the GCE,
$\omega_{g.c.e.}^{+~Boltz}=\omega_{g.c.e.}^{-~Boltz}=1$. When the
net charge density increases the $\omega_{c.e.}^{+~Boltz}$
decreases and tends to 0 at $\mu^*\rightarrow \infty$, while the
$\omega_{c.e.}^{-~Boltz}$ increases and tends to 1.
% at $\mu^*\rightarrow \infty$.
The physical reasons of this are seen from Eq.\,(\ref{nBoltz})
which at $\mu^*\gg 1$ gives: $\rho_+\simeq \rho_Q$ and $\rho_-\ll
\rho_Q $. Therefore, at $\mu^*\gg 1$ an exact charge conservation
in the CE keeps $N_+$ close to its average value $Q$ and makes the
fluctuations of $N_+$  in the CE small. Under the same conditions,
$\langle N_-\rangle_{c.e.}$ is much smaller than $Q$, so that the
fluctuations of $N_-$  are not affected by the CE suppression
effects and they have the Poisson  form, as the GCE. The
difference between $\omega_{c.e.}^{+~Boltz}$ and
$\omega_{c.e.}^{-~Boltz}$, and their dependence on $\mu^*$, are
both the new features of the CE.  The GCE scaled variances in the
Boltzmann approximation are equal, $\omega_{g.c.e.}^{+
Boltz}=\omega_{g.c.e.}^{- Boltz}=1$, and they do not depend on the
chemical potential.

The scaled variances $\omega_{c.e.}^{\pm Bose}$ and
$\omega_{g.c.}^{\pm Fermi}$, given by Eq.~(\ref{omega-alpha-ce}),
for different values of $m^*$ are shown in Figs.~5 and 6 as
functions of $\mu^*$. At $\mu^*=0$ it follows that $\rho_+=\rho_-$
and $v_p^{+~2}=v_p^{-~2}$. From Eq.\,(\ref{omega-alpha-ce}) we
find then for the CE scaled variances,
% (),
%
\eq{ \label{mu0} \omega^{\pm}_{c.e.}(\mu^*=0)~=~
\frac{1}{2}\,\omega^{\pm}_{g.c.e.}(\mu^*=0)~. }
According to Eq.~(\ref{mu0})  the CE scaled variances
 at $\mu^*=0$ are two times smaller than the corresponding scaled
variances in the GCE,  e.g., for massless Bose and Fermi particles
(see Figs.~5 and 6, and compare with
Eqs.~(\ref{m0Bfluc},\ref{m0Ffluc})):
 \eq{
 % \omega_{c.e.}^{\pm Boltz}(\mu^*=0)  \;=\;\frac{1}{2}\,,~~~~
      \omega_{c.e.}^{\pm Bose}(\mu^*=0,~m^*\rightarrow 0)~  =~
      \frac{1}{2}~\frac{\zeta(2)}{\zeta(3)}~ \simeq ~0.684\,,~~~~
      \omega_{c.e.}^{\pm Fermi}(\mu^*=m^*=0)~  =
      ~\frac{1}{3}~ \frac{\zeta(2)}{\zeta(3)}~ \simeq ~0.456\,. \label{omegaBFce}
}

We study now the CE scaled variances at non-zero values of
$\mu^*$. Let us start with $\omega^{+~Bose}_{c.e.}$ (Fig.~5,
left). At $\mu^* \rightarrow m^*$ it has been found that $\sum_p
v_p^{+~2}\rightarrow \infty$ (see Eq.~(\ref{n2bose})), thus it
follows from Eq.~(\ref{omega-alpha-ce}):
\eq{\label{omegaBpce}
\omega_{c.e.}^{+~Bose}(\mu^*=m^*)~=~\frac{\sum_pv_p^{-~2}}{V\,\rho_+^{Bose}}~=~
\omega_{g.c.e.}^{-~Bose}(\mu^*=m^*)~\times~\frac{\rho_-^{Bose}(\mu^*=m^*)}{\rho_+^{Bose}(\mu^*=m^*)}~.
 }
The first factor  on the right hand side of Eq.~(\ref{omegaBpce}),
$\omega_{g.c.e.}^{-~Bose}(\mu^*=m^*)$, reaches its maximum,
$\zeta(2)/\zeta(3)\simeq 1.368$ (\ref{m0Bfluc}), at
$\mu^*=m^*\rightarrow 0$ (see Fig.~4, right). When
$\mu^*=m^*\rightarrow 0$, the second factor  on the right hand
side of Eq.~(\ref{omegaBpce}),
$\rho_-^{Bose}(\mu^*=m^*)/\rho_+^{Bose}(\mu^*=m^*)$, also
increases and goes to 1. Therefore,
%a value of $\zeta(2)/\zeta(3)\simeq 1.368$
an upper limit for $\omega_{c.e.}^{+~Bose}$ is reached at $\mu^*
=m^* \rightarrow 0$ (see Fig.~5, left):
\eq{\label{maxBp}
max[\omega_{c.e.}^{+~Bose}(\mu^*,m^*)]~=~\omega_{c.e.}^{-~Bose}(\mu^*=0,~m^*\rightarrow
0)~
%=~\frac{\zeta(2)}{\zeta(3)}~
\simeq~1.368~. }

\begin{figure}[ht!]
\hspace{-0.7cm}
 \epsfig{file=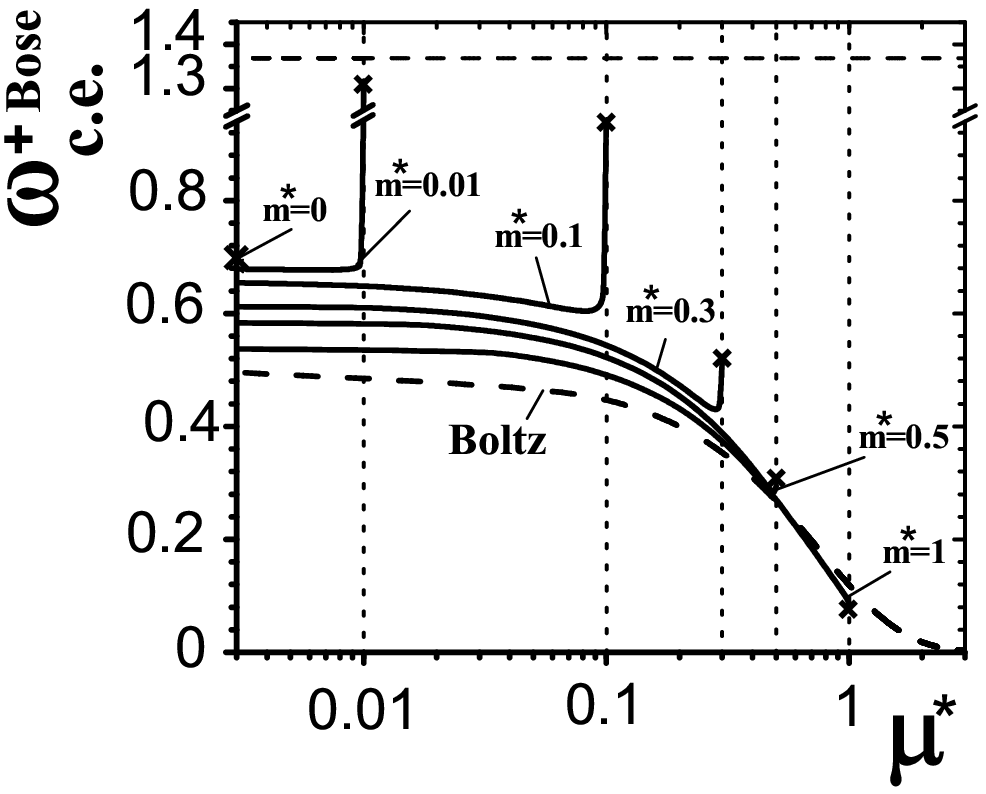,height=6cm,width=8cm}
\hspace{0.7cm}
 \epsfig{file=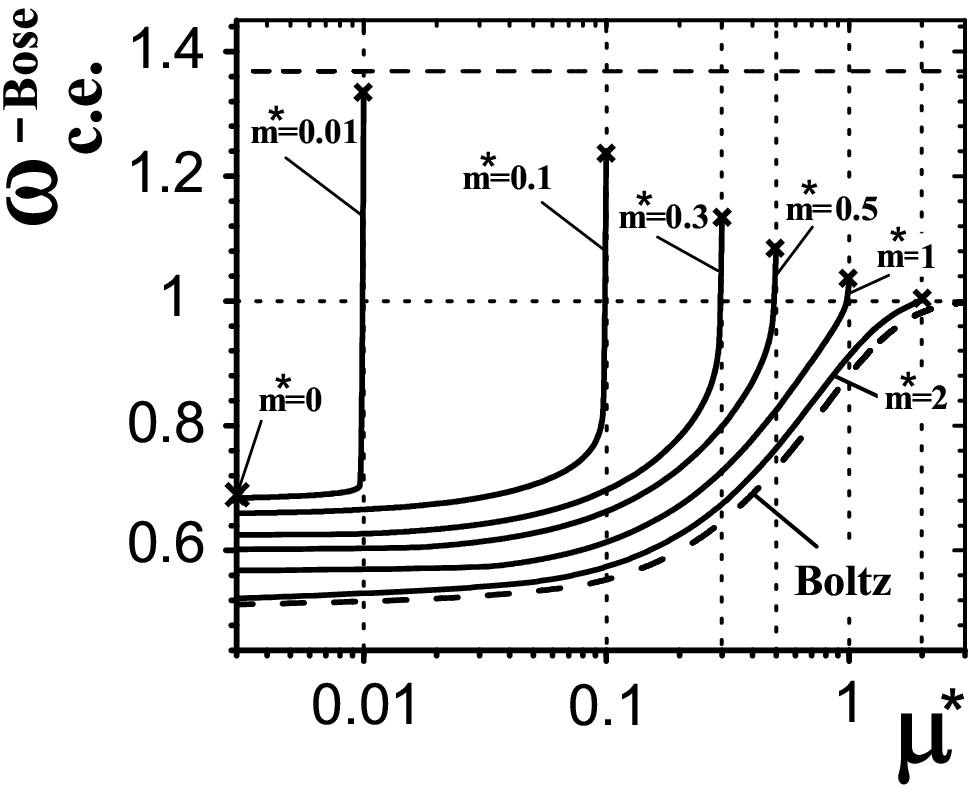,height=6cm,width=8cm}
\caption{ The scaled variances $\omega_{g.c.e.}^{+~Bose}$, left,
and $\omega_{g.c.e.}^{-~Bose}$, right, given by
Eq.~(\ref{omega-alpha-ce}),  are shown as functions of $\mu^*$.
The solid lines present $\omega_{g.c.e.}^{\pm Bose}$ at
$m^*=0.01,~ 0.1,~ 0.3,~ 0.5,~ 1,~ 2,~ 3$.   The vertical dotted
lines $\mu^*=m^*$ demonstrate the restriction $\mu^*\le m^*$ in
the Bose gas.
% (for massless case $m^*=0$ the only value $\mu^*=0$ is
%permitted in the Bose gas).
The dashed horizontal line presents a value of
$\zeta(2)/\zeta(3)\simeq 1.368$ which is an upper limit for
$\omega_{c.e.}^{\pm~Bose}$ reached at $\mu^* =m^* \rightarrow 0$
(see Eqs.(\ref{maxBp},\ref{maxBm})). The crosses at $\mu^*=m^*$
correspond to the points of Bose condensation. The crosses at
$\mu^*=0$ correspond to
$\omega_{c.e.}^{\pm~Bose}(\mu^*=0,m^*\rightarrow 0$ given by
Eq.~(\ref{omegaBFce}). The dashed lines correspond to
$\omega_{c.e.}^{+ Boltz}$, left and $\omega_{c.e.}^{- Boltz}$,
right, given by Eq.~(\ref{omega-alpha1}).
 \label{fig5}}
\end{figure}

\vspace{0.5cm}
%\newpage
\begin{figure}[ht!]
\hspace{-0.7cm}
 \epsfig{file=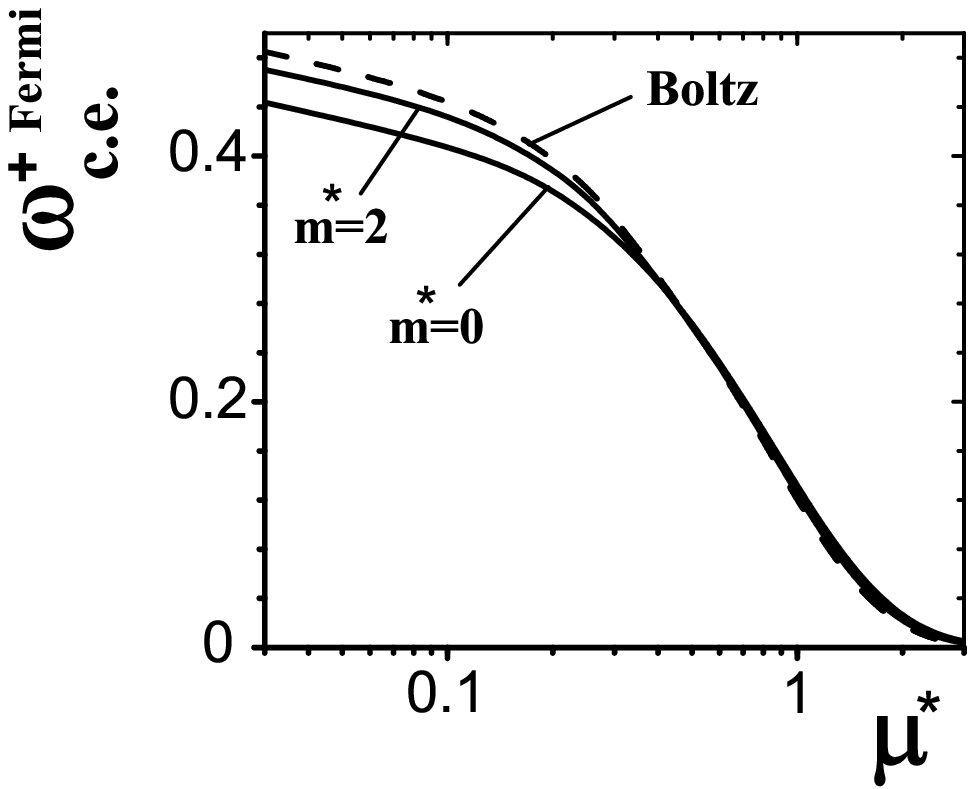,height=6cm,width=8cm}
\hspace{0.7cm}
 \epsfig{file=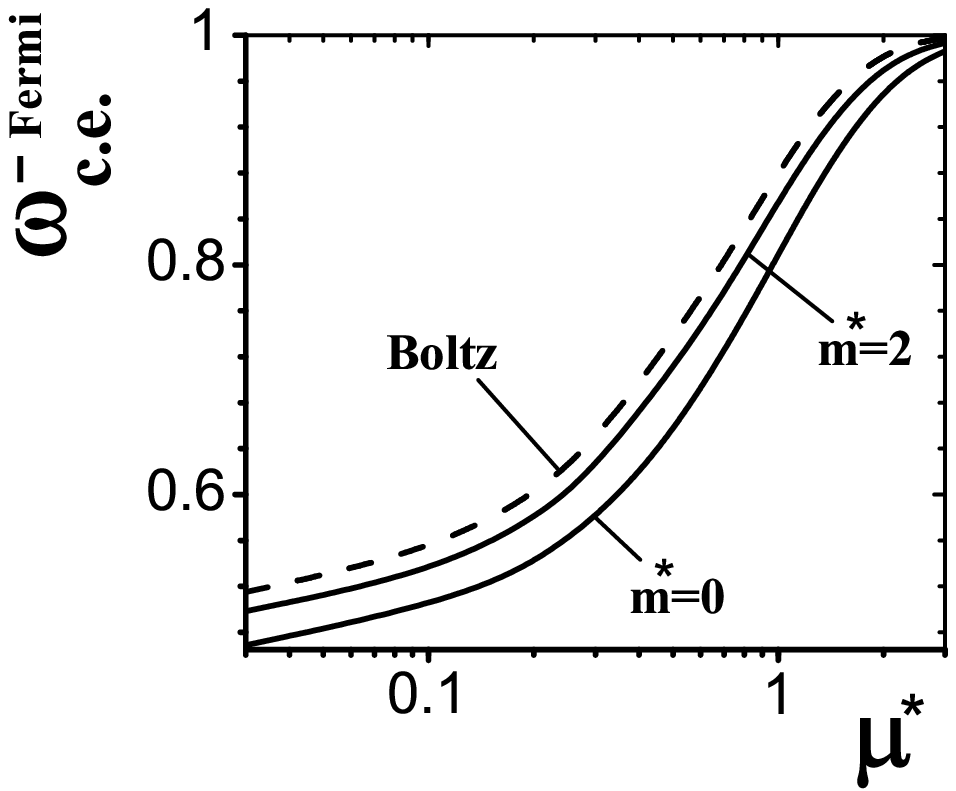,height=6cm,width=8cm}
% \end{figure}
%
% \begin{figure}
\caption{The scaled variances $\omega_{c.e.}^{+ Fermi}$ (left) and
$\omega_{c.e.}^{- Fermi}$ (right) are presented by the solid lines
for $m^*=0$ and $m^*=2$. The dashed lines correspond to
$\omega_{c.e.}^{+ Boltz}$ (left) and $\omega_{c.e.}^{- Boltz}$
(right) given by Eq.~(\ref{omega-alpha1})
 \label{fig6}}
\end{figure}

At $\mu^*=m^*\rightarrow\infty$ one finds
$\omega_{g.c.e.}^{-~Bose}(\mu^*=m^*\rightarrow\infty)\rightarrow
1$ (see Fig.~4, right). Therefore, it follows:
\eq{\label{minBp}
\omega_{c.e.}^{+~Bose}(\mu^*=m^*\rightarrow\infty)~\simeq~
\frac{\rho_-^{Bose}(\mu^*=m^*\rightarrow\infty)}{\rho_+^{Bose}(\mu^*=m^*\rightarrow\infty)}~
\simeq~ \frac{1}{\zeta(3/2)}~\exp\left(-2\mu^*\right)~\simeq~
0.383~\exp\left(-2\mu^*\right)~, }
so that at $\mu^*=m^*\rightarrow\infty$ the scaled variance
$\omega_{c.e.}^{+~Bose}$ goes to zero faster than
$\omega_{c.e.}^{+~Boltz}\simeq \exp(-2\mu^*)$. The Fig.~5 (left)
demonstrates that $\omega_{c.e.}^{+~Bose}(\mu^*=m^*=1)$ is already
smaller than $\omega_{c.e.}^{+~Boltz}(\mu^*=1)$ (Bose
'suppression' !).

For $\omega_{c.e.}^{-~Bose}(\mu^*=m^*)$ (see Fig.~5, right) one
finds from Eq.~(\ref{omega-alpha-ce}):
\eq{\label{omegaBmce}
\omega_{c.e.}^{-~Bose}(\mu^*=m^*)~=~\omega_{g.c.e.}^{-~Bose}(\mu^*=m^*)~,
}
so that
\eq{\label{maxBm}
max[\omega_{c.e.}^{-~Bose}(\mu^*,m^*)]~=
~\omega_{g.c.e.}^{-~Bose}(\mu^*=0,~m^*\rightarrow 0)~
%=~\frac{\zeta(2)}{\zeta(3)}~
\simeq~1.368~. }
From Eqs.~(\ref{omegagce4},\ref{omegaBmce}) it follows,
\eq{\label{omegagBmce1}
\omega^{-~Bose}_{c.e}~\simeq~1~+~\gamma~2^{-3/2}~
\exp\left(-~2m^*\right)~, }
 and $\omega_{c.e.}^{-~Bose}(\mu^*=m^*
\rightarrow\infty)$ goes to 1 from above (see Fig.~4, right).

Now let us turn to the behavior of $\omega_{c.e.}^{+~Fermi}$ and
$\omega_{c.e.}^{-~Fermi}$ (Fig.~6, left and right, respectively).
The variance $\;\sum_{p} v_{p}^{+\,2}\;$ for the Fermi gas
increases as $\mu^{*2}$, whereas  $\;\sum_{p} v_{p}^{-\,2}\;$
decreases exponentially, $\exp(-\mu^*)$, for large chemical
potentials, $\;\mu^{*}\gg 1\;$. Then it follows  from
Eq.~(\ref{omega-alpha-ce}),
\eq{\label{omegaFpce}
\omega_{c.e.}^{+~Fermi}~\simeq ~ \omega_{g.c.e.}^{-~Fermi}~
\times~ \frac{\rho_-^{Fermi}}{\rho_+^{Fermi}}~\simeq~
1~\times~\frac{m^{*2}K_2(m^*)\exp(-\mu^*)}{\mu^{*3}/3} ~, }
for $\mu\rightarrow\infty$. Therefore, $\omega_{c.e.}^{+~Fermi}$
goes to zero  like $\mu^{*-3}\exp(-\mu^*)$ as
$\mu^*\rightarrow\infty$. However, $\omega_{c.e.}^{+~Boltz}\simeq
\exp(-2\mu^*)$, and $\omega_{c.e.}^{+~Fermi}$ becomes larger than
$\omega_{c.e.}^{+~Boltz}$  (Fermi 'enhancement' !) as
$\mu^*\rightarrow\infty$ (see Fig.~6, left). Finally, using
Eqs.~(\ref{RFpm},\ref{Fmgce},\ref{A14}) one  finds for
$\omega_{c.e.}^{-~Fermi}$ as $\mu^*\rightarrow\infty$:
\eq{\label{omegaF1pce}
\omega_{c.e.}^{-~Fermi}~&\simeq ~ \omega_{g.c.e.}^{-~Fermi}~
\times~ \left(1~-~\frac{\sum_p v_p^{-~2}}{\sum_p
v_p^{+~2}}\right)~\simeq~\left[1~-~\frac{K_2(2m^*)}{2K_2(m^*)}~
\exp\left(-~\mu^*\right)\right]~\nonumber
\\
& \times~\left[1~-~\left(\frac{m^*}{\mu^*}\right)^2
K_2(m^*)\exp\left(-\mu^*\right)\right]~.
}
Therefore, $\omega_{c.e.}^{-~Fermi}$ goes to 1 at
$\mu^*\rightarrow\infty$ satisfying  the inequalities (see Fig.~6,
right):
\eq{\omega_{c.e.}^{-~Fermi}~<~
\omega_{g.c.e.}^{-~Fermi}~<~\omega_{c.e.}^{-~Boltz}~. }

%
%%%%%%%%%%%%%%%%%%%%%%%%%%%%%%%%%%%%%%%%%%%%%%%%%%%%%%%%%%%%%%%%%%%%%%%%%%%%%%%%%%%%%%%%
\section{Summary}
%%%%%%%%%%%%%%%%%%%%%%%%%%%%%%%%%%%%%%%%%%%%%%%%%%%%%%%%%%%%%%%%%%%%%%%%%%%%%%%%%%%%%%%%
%
The scaled variances for the particle number fluctuations have
been systematically studied for the Bose and Fermi ideal
relativistic gases.  The calculations have been done in the grand
canonical and  canonical ensembles.  The analysis reveals that in
the limit of large (positive) chemical potential the quantum
effects and effects of the exact charge conservation are absent
for negatively charged particles, so that $\omega^{-~Bose}\simeq
\omega^{-~Fermi}\simeq \omega^{-~Boltz}\simeq 1$ in both the GCE
and CE. However,  the strongest quantum effects  take place in the
limit of large chemical potential for the fluctuations of number
of positively charged particles in the GCE (see Fig.~4):
$\omega^{+~Bose}_{g.c.e.}\rightarrow\infty$  as $\mu^*\rightarrow
m^*$ and $\omega^{+~Fermi}_{g.c.e.}\rightarrow 0$ as
$\mu^*\rightarrow \infty$.
%the scaled variance for positively
%charged Bose particles tends to infinity at $\;\mu\rightarrow
%m\;$, where Bose condensation starts.  In the case of Fermi
%positively charged particles the scaled variance decreases to zero
%$\;\mu \rightarrow\infty$.
% The scaled variance
%for negatively charged particles tends to unity for both bosons
%and fermions in the G.C.E.
On the other hand, just in the limit of large chemical potential
we have found the strongest effects of the exact charge
conservation. The scaled variances $\omega^{+~Bose}_{c.e.}$ and
$\omega^{+~Fermi}_{c.e.}$ (See Figs.~5 and 6) at large $\mu^*$ are
very different from those in the GCE. The Bose and Fermi effects
in the CE are clearly seen at intermediate $\mu^*$, but for
$\mu^*\gg 1$ the effects of the exact charge conservation
dominate: $\omega^{+~Bose}_{c.e.}\simeq
\omega^{+~Boltz}_{c.e.}\rightarrow 0$ at
$\mu^*=m^*\rightarrow\infty$ and $\omega^{+~Fermi}_{c.e.}\simeq
\omega^{+~Boltz}_{c.e.}\rightarrow 0$ at $m\ll
\mu\rightarrow\infty$ .
% The situation is more complicated in the C.E..
%For bosons the scaled variance has a maximum for both positively
%and negatively charged particles at the point $\;\mu^*=m^*\;$, in
%contrast to G.C.E where the scaled variance is infinite at this
%point. This maximum decrease with increasing mass of particles and
%exists for any small but nonzero mass. An another very interesting
%result is that the scaled variance for positively charged
%particles in C.E. changes with increasing chemical potential from
%Bose enhancement to Bose suppression and from Fermi suppression to
%Fermi enhancement  for bosons and fermions correspondingly.
The summary of analytical results for
some limiting values of the scaled variances for
positively and negatively charged particles  
in the GCE and CE are presented in Table 1.

\begin{table}[ht!]
\centering
\begin{tabular}{|c|c|c|c|c|c|c|}
\hline
 && $\mu^*=0,~m^*\to 0$ & $\mu^*=m^*\to0$ & $\mu^*\ll m^*\to\infty$&
 $\mu^*=m^*\to\infty$ & $m^*\ll\mu^*\to\infty$ \\
 %\hline\\
\cline{3-7}
   & $\omega^{+~Boltz}_{g.c.e}$ & 1 & 1& 1 & 1 & 1 \\
%  \textcolor{red}{
Grand
& $\omega^{-~Boltz}_{g.c.e}$ & 1 & 1  & 1& 1 &1 \\
  \cline{2-7}
%  \textcolor{red}{
Canonical
    & $\omega^{+~Bose}_{g.c.e}$ & 1.368 & $\infty$ &1 & $\infty$ & --- \\
% \textcolor{red}{
Ensemble & $\omega^{-~Bose}_{g.c.e}$ & 1.368 &1.368 & 1 & 1  & --- \\
    \cline{2-7}
   & $\omega^{+~Fermi}_{g.c.e}$ & 0.912 & 0.912 & 1 & 0.791 & 0 \\
 &$\omega^{-~Fermi}_{g.c.e}$ & 0.912 & 0.912 & 1&  1& 1\\
  \hline\hline
& $\omega^{+~Boltz}_{c.e}$ & 0.5 & 0.5 & $0.5[1-\tanh(\mu^*)]$ &
$0.5[1-\tanh(\mu^*)]\rightarrow 0$ & $0.5[1-\tanh(\mu^*)]\rightarrow 0$ \\
   & $\omega^{-~Boltz}_{c.e}$ & 0.5 & 0.5 & $0.5[1+\tanh(\mu^*)]$
    & $0.5[1+\tanh(\mu^*)]\rightarrow 1$ &$0.5[1+\tanh(\mu^*)]\rightarrow 1$ \\
    \cline{2-7}
% \textcolor{red}{
Canonical   & $\omega^{+~Bose}_{c.e}$ & 0.684 & 1.368 &$0.5~[1-\tanh(\mu^*)]$& 0 & --- \\
%\textcolor{red}{
Ensemble &$\omega^{-~Bose}_{c.e}$ & 0.684 &1.368 & $0.5~[1+\tanh(\mu^*)]$ & 1 & --- \\
    \cline{2-7}
    &$\omega^{+~Fermi}_{c.e}$ & 0.456 & 0.456 &$0.5~[1-\tanh(\mu^*)]$ & 0 & 0 \\
 &$\omega^{-~Fermi}_{c.e}$ & 0.456 & 0.456& $0.5~[1+\tanh(\mu^*)]$ & 1 & 1 \\
 \hline
\end{tabular}
  \caption{The scaled variances $\omega^+$ and $\omega^-$ for different
  statistics in the GCE and CE.
  %For massless Bose particles
  %nonzero values of the chemical potentials are forbidden,
  % For the Bose particles
%and there are discontinuities between the limits of
%$\omega^{+~Bose}_{g.c.e.}$, $\omega^{+~Bose}_{c.e.}$, and
%$\omega^{-~Bose}_{c.e.}$ at $\mu^*=m^*\rightarrow 0$ and their values
%at $\mu^*=m^*=0$.
The values of $\mu^*>m^*$ are forbidden
  in the Bose gas.
  }\label{t-1}
\end{table}

%
%\newpage
 \begin{acknowledgments}
 % \vspace{0.3cm} \noindent{ \bf Acknowledgements.}
We would like to thank F.~Becattini, A.I.~Bugrij, M.~Ga\'zdzicki,
W.~Greiner, V.P.~Gusynin,
% A.~Ker\"anen,
A.P.~Kostyuk, I.N.~Mishustin, St.~Mr\'owczy\'nski, 
Y.M.~Sinyukov, H. St\"ocker, and O.S.~Zozulya
%, L.M.~Satarov
for useful discussions and comments. We thank  B.O$'$Leary and
Z.I. Vakhnenko for help in the preparation of the manuscript. The
work was supported by US Civilian Research and Development
Foundation (CRDF) Cooperative Grants Program, Project Agreement
UKP1-2613-KV-04.
\end{acknowledgments}

\appendix

\section{}
%Asymptotic behavior
An integral representation of the modified Hankel function $K_2$
has the form \cite{AS}:
 \eq{K_2\left(n\,m^*\right)~=~n~m^{*\,-2}~
 \int_0^{\infty}x^2dx~\exp\left(-n\,\sqrt{m^{*2}+x^2}\right)~.
 \label{a1}
 }
The asymptotic behavior of the $K_2$ function at large and small
arguments is the following:
 \eq{
 K_2(y)\;& \simeq\;\sqrt{\frac{\pi}{2y}}\,\exp(-y)\;,~~~~ y~\gg~
 ~1~; \label{a2} \\
K_2(y)\;& \simeq\;2~y^{-2}\;,~~~~~~ y~\ll~
 ~1~.\label{a3}
 }
Using for $z>0$ the expansions ($\gamma=+1,-1,0$),
 \eq{
 \frac{1}{\exp(z)~-~ \gamma} \;=\;
 \sum_{n=1}^{\infty}\gamma^{n-1}\exp(-n\,z)\;,~~~~~
 \frac{1}{\left(\exp(z)~- ~\gamma\right)^2} \;=\;
 \sum_{n=1}^{\infty}\gamma^{n-1}\,n\;\exp[-(n+1)\,z]\;,
 \label{a4}}
one finds at $\mu^*\le m^*$:
 \eq{
 \int_0^{\infty}\frac{x^2dx}{\exp\left[\left(\sqrt{m^{*2}+x^2}\mp\mu^*\right)\right]
~ - ~\gamma}
 &\;=\;
m^{*2}~\sum_{n=1}^{\infty}\frac{\gamma^{n-1}}{n}\,K_2\left(n\,m^*\right)
 \exp\left(\pm n\mu^*\right)\;, \label{a5}
\\
 \int_0^{\infty}\frac{x^2dx}{\left[\;\exp\left[\left(\sqrt{m^{*2}+x^2}\mp\mu\right)\right]
 ~-~\gamma\;\right]^2}
 &\;=\;
 m^{*2}~\sum_{n=1}^{\infty}\frac{\gamma^{n-1}~n}{n+1}K_2\left[(n+1)\,m^*\right]
 \exp\left[\pm (n+1)\mu^*\right]\;.\label{a6}
 }
The polylogarithm function (or Jonqui\`{e}re's function) is
defined as \cite{PBM}
\eq{Li_k(z)~=~
 \sum_{n=1}^{\infty}~ \frac{z^n}{n^k}~.
\label{a7}
 }
For $\;z=1\;$ it  equals the Riemann zeta function,
\eq{ \zeta(k)~=~\sum_{n=1}^{\infty}~\frac{1}{n^k}~. \label{a8}
}
Some special values of the zeta function used in the paper are:
\eq{ \zeta\left(\frac{3}{2}\right)~\simeq~2.612~,~~~~ \zeta\left(2
\right)~=~\frac{\pi^2}{6}~\simeq~1.645,~~~~ \zeta\left(3
\right)~\simeq~1.202,~~~~
\zeta\left(4\right)\;=\;\frac{\pi^4}{90}~\simeq~1.082~.\label{a9}
}

\section{}
%Fermi gas at $\mu^*\gg m^*$.
To obtain the asymptotic expansion of $\rho_+^{Fermi}$ at
$\mu^*\gg m^*$ one calculates making the variable substitutions
and integrating by parts:
 \eq{
 & \int_0^{\infty}\frac{x^2dx}{\exp\left(\sqrt{m^{*2}+x^2}-\mu^*\right)+ 1}
 = \int_{m^*}^{\infty}
        \frac{\sqrt{\epsilon^2-m^{*2}}~\epsilon~ d\epsilon}
        {\exp\left(\epsilon-\mu^*\right)+1}
=\frac{1}{3}
\int_{m^*-\mu^*}^{\infty}dy~[(y+\mu^*)^2-m^{*2}]^{3/2}
~\frac{\exp(y)}{[\exp(y)+1]^2}
%  \equiv~\int_{m*}^{\infty}
%        \frac{f(\varepsilon)\;d\varepsilon}
%        {\exp\left(\varepsilon-\mu^*\right)~+~1}
        \nonumber \\
&~\equiv~\frac{1}{3} \int_{m^*-\mu^*}^{\infty}~dy~f(y)
~\frac{\exp(y)}{[\exp(y)+1]^2}~.
 \label{A1} }
The function $\exp(y)[\exp(y)+1]^{-2}$ has a maximum at $y=0$ and
decreases exponentially at $y\rightarrow\pm\infty$. Expanding the
function $f(y)$ in a  Taylor series at $y=0$ and extending the
lower limit of the $y$-integral in Eq.~(\ref{A1})  to $-\infty$
(this adds only exponentially small term proportional to
$\exp(-\mu^*)$) one finds an asymptotic expansion at $\mu^*\gg
m^*$:
 \eq{
\frac{1}{3} \int_{m^*-\mu^*}^{\infty}~dy~f(y)
~\frac{\exp(y)}{[\exp(y)+1]^2}~ \simeq~ \frac{1}{3}
\int_{-\infty}^{\infty}~dy~\left[f(0)~+~f^{\prime}(0)~y~+~\frac{1}{2}~
f^{\prime\prime}(0)~y^2~+~\ldots~\right]
~\frac{\exp(y)}{[\exp(y)+1]^2}.
 \label{A2} }
It follows for $f(0)$ and its derivatives
\eq{
f(0)~=~(\mu^{*~2}~-~m^{*~2})^{3/2}~,~~~~f^{\prime}(0)~=~
3\mu^*~(\mu^{*~2}~-~m^{*~2})^{1/2}~,~~~~ f^{\prime\prime}(0)~=~
3~\frac{2\mu^{*~2}~-~m^{*~2}}{(\mu^{*~2}~-~m^{*~2})^{1/2}}\;,\label{A3}
}
%\;+\;
% \frac{\pi^2}{6}\,f'(\mu^*)~+~\ldots~ \simeq~\frac{1}{3}\,
% \left(\mu^{*2}-m^{*2}\right)^{3/2}
% \;+\;
% \frac{\pi^2}{6}\,\frac{2\mu^{*2}-m^{*2}}{\sqrt{\mu^{*2}-m^{*2}}}~+~\ldots~.
% }
 so that $f^{(n)}(0)\propto (\mu^*)^{3-n}$ as
$\mu^*\rightarrow\infty$. The remaining $y$-integrals are equal to
\cite{PBM}:
 \eq{\label{A4}
 I_0~&=~\int_{-\infty}^{\infty}dy~\frac{\exp(y)}{[\exp(y)~+~1]^2}~=~1~,
 \\
  I_n\;&=\;\int_{-\infty}^{\infty}~dy~y^n~\frac{\exp(y)}{[\exp(y)~+~1]^2}~=~
 2~
 n!~\sum_{k=1}^{\infty}\frac{(-1)^{k+1}}{k^n}~=~2~n!~(1~-~2^{-n+1})~\zeta(n)~,
 \label{A5}}
 for even $n=2l$, and $I_n=0$ for odd $n=2l-1$, ($l=1,2,3,\ldots$).
One finally obtains:
\eq{\label{A6}
&
\int_0^{\infty}\frac{x^2dx}{\exp\left(\sqrt{m^{*2}+x^2}-\mu^*\right)+
1}~\simeq~\frac{1}{3}~\left[
f(0)~I_0~+~\frac{1}{2}\;f^{\prime\prime}(0)~I_2~+~\ldots~\right]
\nonumber\\& =~ \frac{1}{3}~(\mu^{*~2}~-~m^{*~2})^{3/2}~...+~
\frac{2\mu^{*~2}~-~m^{*~2}}{(\mu^{*~2}~-~m^{*~2})^{1/2}}~
\zeta(2)~+~\ldots
 ~\simeq\;
 \frac{1}{3}\;\mu^{*\,3}+
 \left(\frac{\pi^2}{3}-\frac{m^{*\,2}}{2}\right)\mu^*+\ldots~. }
%
%with $\zeta(2)=\pi^2/6$.
%

To find $\omega^{+~Fermi}$ at $\mu^*\gg m^*$ one needs to
calculate integral (\ref{n2}) for $\alpha=-\gamma =1$ (note that
$\omega^{-Fermi}\simeq \omega^{-Boltz}=1$ in this limit). Similar
to Eqs.~(\ref{A1},\ref{A2}) one finds:
   \eq{
% \label{n2}
& \int_0^{\infty}\frac{x^2 dx}{\left[\;\exp
\left(\sqrt{x^{2}+m^{*\,2}}~+ \mu^*
       \right) ~+~ 1\;\right]^2}~\nonumber \\
& \simeq~ \frac{2}{3}
\int_{-\infty}^{\infty}~dy~\left[f(y_0)~+~f^{\prime}(y_0)~(y~-y_0)~+~\frac{1}{2}~
f^{\prime\prime}(y_0)~(y~-~y_0)^2~+~\ldots~\right]
~\frac{\exp(y)}{[\exp(y)+1]^3}~,  \label{A7} }
where $y_0=-\ln 2$ is the point of maximum for the function
$\exp(y)[\exp(y)+1]^{-3}$. The $y$-integrals in Eq.~(\ref{A7}) are
equal to \cite{GR}:
\eq{\label{A8}
A_0~&=~\int_{-\infty}^{\infty}dy~\frac{\exp(y)}{[\exp(y)~+~1]^3}~=~\frac{1}{2}~,
 \\
 A_1\;&=\;\int_{-\infty}^{\infty}~dy~\frac{y~\exp(y)}{[\exp(y)~+~1]^3}~=~
 -\frac{1}{2}~,\label{A9} \\
A_2\;&=\;\int_{-\infty}^{\infty}~dy~\frac{y^2~\exp(y)}{[\exp(y)~+~1]^3}~=~
 \frac{\pi^2}{6}\label{A10}~.
 }
One finds: \eq{\label{A11}
&
\int_0^{\infty}\frac{x^2dx}{\left[\exp\left(\sqrt{m^{*2}+x^2}-\mu^*\right)+
1\right]^2}\nonumber
\\
&\simeq~\frac{2}{3}~\left[f(y_0)~A_0~+~f^{\prime}(y_0)~(A_1~-~y_0~A_0)~+~\frac{1}{2}~
f^{\prime\prime}(y_0)~(A_2~-~2y_0~A_1~+~y_0^2~A_0)~+~\ldots~\right]
 \nonumber \\
 &\simeq\;
 \frac{1}{3}\;\mu^{*\,3}-\mu^{*\,2}+
 \left(\frac{\pi^2}{3}-\frac{m^{*\,2}}{2}\right)\mu^*+\ldots
}
Thus for Fermi particles at $\mu^*\rightarrow\infty$ the following
expansions are obtained:
 \eq{
 \sum_p\langle n_p^{+}\rangle_{g.ce.}
& \;\simeq\;\frac{g\,V\,T^3}{2\pi^2}
 \left[\frac{1}{3}\;\mu^{*\,3}+
 \left(\frac{\pi^2}{3}-\frac{m^{*\,2}}{2}\right)\mu^*\right]\;,\label{A12}
 \\
 \sum_p\langle n_p^{+\,2}\rangle_{g.c.e.}
&\;\simeq\;\frac{g\,V\,T^3}{2\pi^2}
 \left[\frac{1}{3}\;\mu^{*\,3}-\mu^{*\,2}+
 \left(\frac{\pi^2}{3}-\frac{m^{*\,2}}{2}\right)\mu^*\right]
 \;=\; \sum_p\langle n_p^{+}\rangle_{g.c.e.} -
 \frac{g\,V\,T^3}{2\pi^2}\;\mu^{*\,2}\;,\label{A13}\\
%
%
%
%
%\eq{
\sum_{p}v_{p}^{+\,2} & \;\equiv\;
 \sum_p\langle n_p^{+}\rangle_{g.c.e.}
 \;-\; \sum_p\langle n_p^{+2}\rangle_{g.c.e.}
 \;\simeq\;\frac{g\,V\,T^3}{2\pi^2}\;\mu^{*\,2}~.
\label{A14} }

\end{document}